\documentclass[10pt,conference]{llncs2e/llncs}

\usepackage{graphicx}
\usepackage[ruled,vlined,linesnumbered]{algorithm2e}
\usepackage{amsmath}
\usepackage{amsfonts}
\usepackage{amssymb}
\usepackage{amsthm}
\usepackage{complexity}
\usepackage{mathtools}
\usepackage{braket}
\usepackage{dirtytalk}
\usepackage[inline]{enumitem}
\usepackage{mathtools}
\usepackage{xspace}
\usepackage[colorlinks,citecolor=blue!60!black!70,linkcolor=blue!60!black!70]{hyperref}
\usepackage[nameinlink]{cleveref}
\usepackage{multirow}

\usepackage{scalerel}
\usepackage[hang,flushmargin]{footmisc}
\usepackage{dsfont}
\usepackage[framemethod=tikz]{mdframed}
\usepackage{mfirstuc}
\usepackage{bm}
\usepackage{scalerel}
\usepackage{stmaryrd}
\usepackage{filecontents}

\usepackage[font={it}]{caption}
\usepackage{subcaption}
\usepackage[section]{placeins} 
\usepackage{tikz}
    \usetikzlibrary{automata, shapes, positioning, arrows,patterns}
\usepackage{xparse}
\usepackage{array}
    \newcolumntype{L}[1]{>{\raggedright\let\newline\\\arraybackslash\hspace{0pt}}m{#1}}
    \newcolumntype{C}[1]{>{\centering\let\newline\\\arraybackslash\hspace{0pt}}m{#1}}
    \newcolumntype{R}[1]{>{\raggedleft\let\newline\\\arraybackslash\hspace{0pt}}m{#1}}

\usepackage{multicol}
\usepackage{parskip}

\usepackage[textsize=tiny]{todonotes}

\usepackage{ifthen}

\usepackage{grffile}

\usepackage{cprotect}
\usepackage{booktabs}

\usepackage{xspace}
\usepackage{multicol}
\usepackage{varwidth}
\usepackage{wrapfig}

\renewcommand\paragraph[1]{\vspace{.0ex}\par\noindent\textbf{#1}.\;\;}

\let\phi=\varphi 
\let\epsilon=\varepsilon

\renewcommand\implies{\Rightarrow}

\DontPrintSemicolon
\SetKwFunction{Fpropa}{propagate}
\SetKwFunction{Fhif}{highest\_inductive\_frame}
\SetKwFunction{Fgen}{generalize}
\SetKwFunction{Fremove}{remove\_cube}
\SetKwFunction{Fmic}{generalize}
\SetKwFunction{Fdown}{down}
\SetKwFunction{Fblock}{block}
\SetKwFunction{Ftop}{top}
\SetKwFunction{Fmax}{max}
\SetKwFunction{Fstrenghten}{strengthen}
\SetKwFunction{Fpdr}{refine}
\SetKwFunction{Fstrategy}{pdr\_strategy}
\SetKwFunction{Fipdrmain}{pdr-main}
\SetKwFunction{Fipdrconstrain}{ipdr\_constrain}
\SetKwFunction{Fipdrrelax}{ipdr-relax}
\SetKwFunction{Fipdrconstrainloop}{constrain}
\SetKwFunction{Fipdrrelaxloop}{relax}
\SetKwFunction{Fsubsume}{remove\_subsumed}

\SetKwFunction{Fpeter}{peterson}
\SetKwFunction{Fgraph}{graph-search}

\SetKw{KwST}{\ s.t.\ }
\SetKwFor{Loop}{loop}{}{end}

\SetKwInput{In}{In\hspace{2.8mm}}
\SetKwInput{Out}{Out}
\SetKwInOut{Pre}{Pre}
\SetKwInOut{Post}{Post}
\SetKwInOut{Vars}{Vars}

\newcommand\defmath[2]{\newcommand#1{\ensuremath{#2}\xspace}}
\newcommand\concept[1]{\textit{#1}}

\newcolumntype{H}{>{\setbox0=\hbox\bgroup}c<{\egroup}@{}}

\providecommand{\tuple}[1]{\ensuremath{\left( #1 \right)}}
\providecommand{\set}[1]{\ensuremath{\left\lbrace #1 \right\rbrace}}

\providecommand{\sizeof}[1]{\ensuremath{\left\vert{#1}\right\vert}}

\newcommand{\defn}{\,\triangleq\,}

\newcommand{\turner}[3][10em]{
  \rlap{\rotatebox{#2}{\begin{varwidth}[t]{3.8cm}#3\end{varwidth}}}%
}

\input{lib/algorithm2e-ext.sty}
\input{lib/commands.sty}
\input{lib/envs-refs.sty}
\input{lib/experiments.sty}
\input{lib/figures.sty}
\input{lib/formula.sty}
\input{lib/text.sty}

\title{Incremental Property Directed Reachability}

\author{
Max Blankestijn,  {Alfons Laarman}
}

\institute{
Leiden University, Leiden Institute for Advanced Computer Science
\email{max@blankestijn.nl}, 
\email{a.w.laarman@liacs.leidenuniv.nl}
}

\newcommand\cull[2]{#2}

\begin{document}

\maketitle

\begin{abstract}
Property Directed Reachability (\pdr) is a widely used technique for formal verification of hardware and software systems. This paper presents an incremental version of \pdr (\ipdr), which enables the automatic verification of system instances of incremental complexity. The proposed algorithm leverages the concept of incremental SAT solvers to reuse verification results from previously verified system instances, thereby accelerating the verification process. The new algorithm supports both incremental constraining and relaxing; i.e., starting from an over-constrained instance that is gradually relaxed.

To validate the effectiveness of the proposed algorithm, we implemented \ipdr  and experimentally evaluate it on two different problem domains. First, we consider a circuit pebbling problem, where the number of pebbles is both constrained and relaxed. Second, we explore parallel program instances, progressively increasing the allowed number of interleavings. The experimental results demonstrate significant performance improvements compared to Z3's \pdr implementation \spacer. Experiments also show that the incremental approach succeeds in reusing a substantial amount of clauses between instances, for both the constraining and relaxing algorithm.

\end{abstract}

\section{Introduction}
\label{sec:introduction}

	Symbolic model checking based on satisfiability has revolutionized automated verification.
	Initially, symbolic model checkers were based on (binary) decision diagrams~\cite{bryant86,mcmillan}. While they enabled the study of large software and hardware systems~\cite{burch-clarke-mcmillan-dill-hwang,coudert2003unified,clarke2018handbook}, they were inevitably limited by memory constraints because decision diagrams represent all satisfying assignments explicitly.
	\textit{Bounded Model Checking} (BMC)~\cite{biere1999symbolic,biere2009bounded}
	alleviated the need for decision diagrams by encoding the
	behavior of a system directly into propositional logic and using satisfiability~\cite{biere2009handbook}, in a way similar to the reductions provided by Cook~\cite{cook} and Levin~\cite{levin1973universal} much earlier (who could have foreseen this future application of the theory?).
	BMC, in turn, is limited by the depth of the system under verification, since the encoding explicitly `unrolls' the transition relation for each time step of the computation
	and each unrolling requires another copy of the state variables.
	The introduction of the IC3 algorithm~\cite{bradley2011sat,bradley2007checking}, later known as Property Directed Reachability (\pdr)~\cite{efficient-pdr,Welp2013,Hoder2012}, circumvents this
    unrolling by using small SAT solver queries to incrementally construct an inductive invariant from the property.

	This work is inspired by the success of BMC and other verification methods based on \concept{incremental SAT solving}~\cite{sat-ass,sat-ass2,bmc-graph,barrett2009handbook}.
	In order to reduce the search space, modern SAT solvers learn new clauses, further constraining the original problem, from contradictions arising during the search for satisfying assignments~\cite{silva1997grasp,barrett2009handbook}. BMC can exploit the power of clause learning by incrementally increasing the hardness of the problem instance, while retaining learned clauses from `easier' instances~\cite{bmc-graph}. A natural parameter here is the unrolling depth: Incremental BMC increases the unrolling bound to generate a new problem instance. This approach has shown to yield multiple orders of magnitude runtime improvements~\cite{hwmcc07}, which often translates into  unrollings that are multiple times longer.
		
	A natural question is whether \pdr can also benefit from incremental SAT solving. However, since \pdr does not unroll the transition relation, we need new parameters to gradually increase the hardness of instances and exploit incremental solvers. Moreover, the standard \pdr algorithm requires an extension to reuse information learned in previous runs. We provide both these parameters and a new incremental \pdr algorithm.
	
	For instance, an increasing parameter to consider, other than the unrolling depth, is the number of parallel threads in a system. However, it is not always clear how a system with fewer threads relates to a larger one, since it is not necessarily either an over- or under-approximation: Interaction between threads can remove behavior in some systems, while the new thread also introduces new behavior. And the incremental SAT solving requires either a relaxing or a constraining of problem instances. Therefore, we focus here on bounding the number of interleavings in a parallel program. Research has shown that most bugs occur after a limited number of interleavings~\cite{context-switch,Grumberg:2005:PUM:1040305.1040316}, so an incremental \pdr algorithm can exploit this parameter by solving a `relaxed' instance bounded to $\ell+1$ interleavings by reusing the previous results (learned clauses in the SAT solver) from solving an instance of $\ell$ interleavings.
	
	Another interesting application of incremental \pdr is for optimization problems. An example is the \PSPACE-complete circuit pebbling problem~\cite{lingas1978pspace}, used to study memory bounds. It asks whether a circuit, viewed as a graph, can be `pebbled' using $p$ pebbles and the optimization problem is to find the lowest~$p$. These pebbles model the memory required to store signals at the internal wires of the circuit (the working memory): An outgoing wire of a gate can be `pebbled' if its incoming wires are and pebbles can always be removed (memory erasure). In the reversible pebbling problem~\cite{reversible-pebbling}, pebbles can only be removed if all incoming wires are pebbled, is relevant for reversible and quantum computations. 
	An incremental \pdr algorithm can potentially solve a `relaxed' instance with $p+1$ pebbles faster by reusing the results from a previous run on $p$ pebbles. Moreover, it could approximate the number of pebbles from above by solving a `constrained' instance with $p-1$ pebbles, by reusing the same previous results (we could start for example with $p$ equal the number of gates in the circuit, which is always enough to successfully pebble it).
	
	In \autoref{sec:ipdr}, we introduce an incremental \pdr (\ipdr) algorithm that can both handle relaxing, as well as constraining systems. It runs an adapted \pdr algorithm multiple times on instances of increasing (or decreasing) `constraintness'. We show how the \pdr algorithm can be adapted to reuse the internal state from a previous run to achieve this. Moreover, \ipdr can combine both constraining and relaxing approaches in a binary search strategy in order to solve optimization problems, such as the pebbling problem.

	We emphasize here that the incremental approach of \ipdr is orthogonal to the incrementality already found in the original \pdr algorithm:
	Its original name IC3 comes from (IC)$^3$, which stands for: 
	``Incremental Construction of Inductive Clauses for Indubitable Correctness.''
	This refers to the internals of the \pdr algorithm which maintain a sequence of increasingly constrained formulas (clause sets) which ultimately converge to the desired inductive invariant. This sequence is also extended \emph{incrementally}.
	However, \ipdr, in addition, incrementally grows a sequence of problem instances that are increasingly (or decreasingly) constrained. This sequence exists in between runs of the adapted \pdr algorithm.
	Both approaches exploit incremental SAT solver capabilities: Or rather, \ipdr uses incremental SAT solving in two different and orthogonal ways: inside \pdr runs and in between \pdr runs on incremental instances.

	In \autoref{sec:experiments}, we give an open source implementation of an \ipdr-based model~checker.
	We experimentally compare \ipdr with Z3's \pdr implementation \spacer~\cite{z3,spacer} and in our own \pdr implementation. From the results, we draw two separate conclusions: 1) for the relaxing of interleavings, \ipdr can reuse a large amount of information between increments, which gives roughly a 30\% performance gain with respect to our own naive \pdr implementation, while outperforming  \spacer as well, and 2) for the pebbling problems,  when gradually constraining the system, \ipdr again reuses many clauses between incremental instances and can achieve performance gains of around 50\% with respect to \spacer and our naive \pdr implementation, but relaxing and binary search (using both relaxing and constraining) do not provide further improvements.

\section{Preliminaries}
\label{sec:prelim}

\defmath\bool{\mathbb B}

Given a set of Boolean variables $\vars = \set{x_1,\ x_2,\ \ldots,\ x_n}$,
a propositional formula ${F}(\vars)$ represents a function
$ F\colon \bool^n \to \bool$ where $\bool \defn \set{0,1}$. A literal is a variable $x_i$ or its negation $\neg x_i$ (also written as $\overline{x_i}$). A clause is a disjunction of literals and a cube a conjunction of literals. A formula in conjunctive normal form (CNF) is a conjunction of clauses and a formula in disjunctive normal form (CNF) is a disjunction of cubes.
A (truth) assignment $v$ is a function $v \colon \vars \to \bool$, which can also be expressed as a cube $\bigwedge_{x_i \in \vars} x_i \Leftrightarrow v(x_i)$.
Vice versa, we can think of the formula $ F$ as a set of satisfying assignments
$\set{v \in \bool^X \mid  F(v) = 1 }$.
We often use this duality to interpret a formula $F(X)$ as both a set of system states (satisfying assignments) and its CNF description (which we may explicitly denote with \form F).

In symbolic model checking, a formula can represent the current set of states of the system under analysis (each satisfying assignment of the constraint formula represents one state). To reason over the next system states, i.e., the system states after the system performs a transition, we use primed variables, e.g., ${F}(\pvars)$, or more concisely: ${F}^\prime$.
So ${F}^\prime$ is obtained by taking every variable $x_i$ in ${F}$, and replacing it with the corresponding $x^\prime_i$. E.g., if ${F} = (a \land b) \lor (a \land \neg c)$ then  ${F}^\prime = (a^\prime \land b^\prime) \lor (a^\prime \land \neg c^\prime)$.
A {symbolic transition system} (\autoref{def:ts}) describes the behavior of  discrete systems in Boolean logic over a set of Boolean variables~
\vars.    \autoref{ex:sokoban} shows how to encode a simple system as an STS.

    \begin{definition}[Symbolic Transition System (STS)] \label{def:ts}
    A symbolic  transition system is a tuple $\ts \defeq (X,\  I,\ \trans)$ where:
    \begin{itemize}
    	\item $S \defeq \bool^\vars$ is the set of all system states defined over Boolean variables $\vars = \set{x_1,\ x_2,\ \ldots,\ x_n}$ (wlog). 
        A system state ${s}\in S$ is an assignment $\vars \to \bool$.
        
    	\item $I \subseteq S$ is a finite set of initial states of the system.
    	
    	\item $\trans \subseteq S \times S^\prime$ is the transition relation. Where $S^\prime$ represents the states $S$ in the next state of the system. If there exists a pair of states $( p, q) \in \trans$, this means that the system can go from state $ p$ to state $ q$ in a single step.
    \end{itemize}
    \end{definition}
    
\begin{figure}[t] 
        \centering
        \vspace{-1em}
        \vspace{-1em}
        \begin{subfigure}{0.4\textwidth}
            \begin{tikzpicture}
                \node[draw, circle] at (0,0) (a) [initial,initial text={}] {$a$};
                \node[draw, circle] at (2,0) (b) {$b$};
                \node[draw, circle] at (2,-1) (c) {$c$};
                \node[draw, circle] at (0,-1) (d) {$d$};
                
                \draw[->] (a) edge[bend left=30] (b);
                \draw[->] (a) edge (d);
                \draw[->] (b) edge[bend left=30] (a);
                \draw[->] (b) edge (c);
            \end{tikzpicture}
        \caption{STS $\trans$}\label{a}
        \end{subfigure}
        \begin{subfigure}{0.4\textwidth}
            \begin{tikzpicture}
                \node[draw, circle] at (0,0) (a) [initial,initial text={}] {$a$};
                \node[draw, circle] at (2,0) (b) {$b$};
                \node[draw, circle] at (2,-1) (c) {$c$};
                \node[draw, circle] at (0,-1) (d) {$d$};
                
                \draw[->] (a) edge[bend left=0] (b);
                \draw[->] (b) edge (c);
            \end{tikzpicture}
        \caption{STS $\Gamma$}\label{b}
        \end{subfigure}
        \caption{STS $\Delta$ relaxes $\Gamma$, since $\Delta \sqsupset \Gamma$.
        		STS $\Gamma$ constrains $\Delta$, since $\Gamma \sqsubset \Delta$.}
        \label{fig:tsts}
\end{figure}
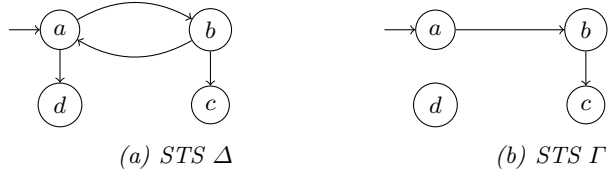

    \begin{example}\label{ex:sokoban}
    We construct an STS $(X,\  I,\ \trans)$ for the system in \autoref{b}. First, we define its states over variables $X = \set{x_1, x_2}$, denoting $00$ for $x_1=x_2=0$, etc. We encode its states as $a = 00$, $b = 01$, $c = 10$ and $d = 11$. The initial states can de encoded as $I(x_1,x_2) = \compl{x_1}\land \compl{x_2}$. Finally, the transition relation is encoded as: $\Delta(x_1,x_2,x_1',x_2') = \compl{x_1} \land x_1'  \Leftrightarrow x_2 \land x_2'  \Leftrightarrow \compl{x_2}$ or alternatively as
    	$(\compl{x_1} \land \compl{x_2} \land \compl{x_1'} \land x_2') \lor 
    	(\compl{x_1} \land {x_2} \land {x_1'} \land \compl{x_2'})$. 
    	Both encodings can be efficiently transformed to CNF for use in SAT solver queries (see e.g. \cite{tseitin1983complexity,PLAISTED1986293}).
    
A basic model checking task is to show that an STS $(X,\  I,\ \trans)$ satisfies an invariant property $P\subseteq S$, i.e., that no state $s \notin  P $ is reachable from the initial states $I$ using the transitions encoded in $\Delta$.
    
    \end{example}

    \paragraph{Image and Preimage}
Given an STS $(X,\ I,\ \trans)$, the image and preimage under $\trans(X,X')$ can be used to reason over the reachable states of a transition system.
    For sets of states $A \subseteq S$, we define the (pre)image of $A$ under $\trans$ as the states (backwards) reachable from $A$-states in one step as follows.
    \[A.\trans~\defeq~\set{{q} \in S \mid \exists {p} \in A \colon  \trans({p},{q})}\]
    \vspace{-3ex} 
    \[\trans.A~\defeq~\set{{p} \in S \mid \exists {q} \in A \colon \trans({p}, {q}) }\]
Of course, a SAT solver can only query individual states (satisfying assignments) $s \in A, t \in B$ for formulas $A,B$. In practice, however, we will not compute (pre)images, but only whether the (pre)image is contained in a set of states, e.g., $A.\trans~\subseteq B$. This can be done with a SAT solver query $\sat(A \land \trans \land B')$, which returns false iff no state in $A$ can transit to a state in $B$, and otherwise returns an example $s \in A, t \in B$ such that $\trans(s, t) = 1$.

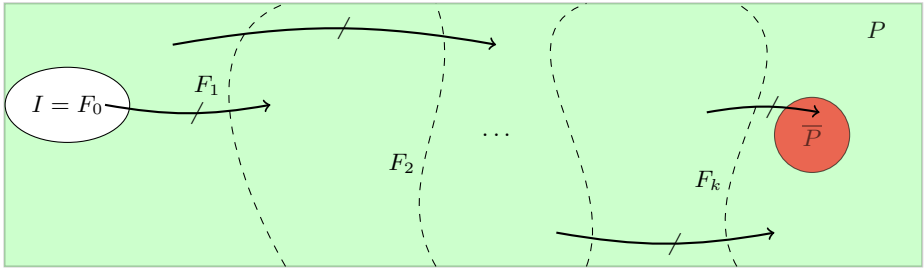
\begin{figure}[b]
	\centering 
	\newlength{\thiswidth}
	\newlength{\thisheight}
	\setlength{\thiswidth}{\textwidth}
	\setlength{\thisheight}{3.5cm}
	\newcommand{\arrowheight}{1.2}

	\begin{tikzpicture}
		
		\tikzset{venn circle/.style={draw, ellipse, minimum width=16cm, minimum height=2.5cm,fill=#1,opacity=0.1}}

		\node [draw, minimum width=\thiswidth, minimum height=\thisheight, opacity=0.2, thick, fill = green] 
			(center) at (-0.38\thiswidth, 0)		{};

		\node [venn circle = white, minimum width=1cm, minimum height=1cm, opacity=1] (I) at (-9.9,.4) {$I=F_0$};
		\node [minimum width=1cm, minimum height=1cm, opacity=0.9] (P) at (0.07\thiswidth, 0.4\thisheight) {$P$};
		\node [venn circle = red, minimum width=1cm, minimum height=1cm, opacity=0.6, text opacity = 1] 
			(nP) at (0,0) {\nP};

		\draw (-8.5, \arrowheight) edge[thick, ->, bend left=10] node{$\not$} (-4.2, \arrowheight);

		\draw (-9.4, .4) edge[thick, ->, bend left=-10] node{$\not$} (-7.2, .4);

		\draw (-1.4, .3) edge[thick, ->, bend left=10] node{$\not$} (0.1, .3);

		\draw (-3.4, -1.3) edge[thick, ->, bend left=-10] node{$\not$} (-0.5, -1.3);

		\draw (-7,-1.75) edge[-,color=black,in=200,out=120,dashed] node[left,pos=.6] {$F_1$} (-7,1.75);
		\draw (-5,-1.75) edge[-,color=black,in=290,out=120,dashed] node[left,pos=.4] {$F_2$} (-5,1.75);
		\draw (-3,-1.75) edge[-,color=black,in=200,out=70,dashed] node[left,pos=.6] {} (-3,1.75);
		\node at (-4.2,0) {$\ldots$};
		\draw (-1,-1.75) edge[-,color=black,in=-30,out=120,dashed] node[left,pos=.3] {$F_k$} (-1,1.75);
		
	\end{tikzpicture} 
	\caption{
		The box represents all states $S = \set{0,1}^X = 1$. The candidates $F_i$ visualized. All $F_i$ for $i \leq k$ are a subset of $P$, each $F_i$
		is a subset of $F_{i+1}$, and there is no transition from a state in $F_i$ to a state in  $F_{i+2}$. 
	}
	\label{ex:frames}
\end{figure}

    \paragraph{Model Checking using the Inductive Invariance Method}
Given an STS $(X,  I,\ \trans)$ and a desired invariant property $P\subseteq S$,
model checking can done by computing all reachable states
$R = I \cup I.\trans \cup I.\trans.\trans \cup \dots$ and showing that $R \subseteq P$.
This is the approach that BMC takes.
Alternatively, \autoref{thm:ii-method} shows that we may also construct an \textit{\ii} (see \autoref{def:ind-inv}) $F \subseteq P$, which contains the initial states:
That is, a strengthening of $P$, which contains $I$ and is also inductive with respect to the transition relation \trans.
E.g., the sets of states $F_1, F_2, \dots, F_k$ in \autoref{ex:frames} prove unreachability of $\overline P$ provided they are inductive.

\begin{definition}[Inductive Invariant] \label{def:ind-inv}
$F \subseteq S$ is an \textit{inductive invariant} for STS $(X, I,\ \trans)$
if  $F.\trans \subseteq F$ (or $ {F} \land \trans \land \neg  {F}^\prime = 0$ as a SAT-solver query).
\end{definition}

\begin{theorem}[Inductive Invariant Method~\cite{iinvariant-method,floyd1993assigning,gribomont1996atomicity}] \label{def:ind-strengthening}\label{thm:ii-method}
A property $P \subseteq S$ is an invariant for STS $(X,  I,\ \trans)$ if and only if there exists   an inductive invariant $F$ such that
$I \subseteq F \text{ and } F \subseteq P$.
\end{theorem}

In particular, $R$ is the strongest possible inductive invariant for an STS $(X, I,\ \trans)$
and all invariant properties $P$ that hold for it.
But in general, an invariant property $P \subseteq S$ is usually not initially inductive, even if it holds for the STS.
For instance, in the case of a mutual exclusion protocol, the property that 
two processes do not end up in the critical section at the same time,
can be violated in states where one process is already in the critical
section and the other is about to enter it. However, in a correct protocol
these states are unreachable. (And this information is contained exactly in the set of reachable states $R$.)

\paragraph{Constraining and Relaxing Symbolic Transition Systems}
	Consider the simple transition systems \textit{(a)} and \textit{(b)} in
	\autoref{fig:tsts}. Here, system \textit{(b)} can be obtained by
	removing transitions from system \textit{(a)}.
	Intuitively speaking, system \textit{(b)} behaves much the same as system
	\textit{(a)} and if \pdr were to collect clauses to describe the reachability
	in \textit{(a)}, that information would seem useful when running \pdr for
	system \textit{(b)}.
	\autoref{def:relaxconstrain} defines relaxing and constraining formally. 
	For instance, if $\Delta$ encodes the transition relation of a pebbling problem with at most $k$ pebbles, then  we have $\Delta \sqsubseteq \Delta\relaxed$ for  $\Delta\relaxed$ encoding the relaxed instance with $k+1$ pebbles.

	\begin{definition}[Constrained and relaxed STSs] \label{def:relaxconstrain}
		An STS $M_1 = (X,  I_1, \trans{}_1)$ is a constrained version of 
		STS $M_2 = (X,  I_2, \trans{}_2)$, denoted $M_1 \sqsubseteq M_2$, iff
		$I_1 \subseteq I_2$ and $\trans{}_1 \subseteq \trans{}_2$.
		Vice versa, we say $M_2$ relaxes $M_1$, or $M_2 \sqsupseteq M_1$.
		
		Consequently, $\sqsubseteq$ is a partial order, and $M_1 = M_2$ iff $M_1 \sqsubseteq M_2$ and  $M_1 \sqsupseteq M_2$.
		We denote $M_1 \sqsubset M_2$  ($M_1 \sqsupset M_2$) when $M_1 \sqsupseteq M_2$ ($M_1 \sqsupseteq M_2$) and $M_1 \neq M_2$.
	\end{definition}


\section{Incremental Property Directed Reachability}
\label{sec:ipdr}

This section introduces a simplified \pdr algorithm.
The full version of \pdr requires intricate interactions
with the SAT solver to attain efficiency (i.e., we omit \concept{generalization} and use set-based notation instead of SAT solver calls).
We emphasize that this simplified \pdr is not efficient as it treats individual states (and not their generalizations); nonetheless, this description suffices to define \ipdr in such a way that it is also compatible with the full version of \pdr (as we discuss in \autoref{sec:pdr}).
For a full description of \pdr, we refer to \cite{understanding-ic3,efficient-pdr,bradley2007checking}.

\autoref{sec:pdr} extends \pdr with an internal state that Incremental \pdr (\ipdr) utilizes to reuse information between \pdr runs.  \ipdr, introduced in \autoref{ss:ipdr}, takes a sequence of constraining (or relaxing) STS instances $M_1, \dots, M_z$ such that $M_{i+1} \sqsubseteq M_{i}$ (or $M_{i+1} \sqsupseteq M_{i}$), solving them one by one with the extended \pdr algorithm, while passing the internal state along to speed up subsequent \pdr runs. 
Using the properties of simplified \pdr (mostly the invariants defined in \autoref{def:frames} maintained by the algorithm that also hold in the full \pdr algorithm), we demonstrate how \ipdr correctly instantiates incremental \pdr runs.

\vspace{-1ex}
\subsection{Extending \pdr with an Internal State}\label{sec:pdr}
\vspace{-1ex}

Given a TST $(X, I, \trans)$ and an invariant property $P \subseteq S$,
constructing an inductive invariant according to \autoref{def:ind-strengthening} is
non-trivial. The \pdr algorithm~\cite{bradley2011sat,efficient-pdr}
approaches this problem by using the concept of
\textit{relative inductivity} (\autoref{def:rel-ind}).
To construct the inductive invariant, \pdr maintains a sequence $F_0, F_1, \dots F_k\subseteq S$ 
of candidate inductive invariants as defined in \autoref{def:frames} and illustrated in \autoref{ex:frames}.
The first candidate $F_0$ is invariably set to the initial states. 
Initially, the algorithm also sets $k=0$ and assures that $I \subseteq P$
so the \Fprop conditions of \autoref{def:frames} are satisfied.

By virtue of the fact that \pdr maintains the \Fprop properties as invariants, a
few observations can be made.
First, the candidates are relatively inductive to their neighbors:
By $\Fprop[2]$, we have $F_{i+1} \cap F_i = F_i$.
Paired with \Fprop[3], this implies that $F_{i+1}$ is
    inductive relative to $F_i$.
It can also be  shown that each candidate $F_i$
over-approximates the set of states reachable within $i$ steps (only states proved unreachable in $i$ steps are blocked in $F_i$).
It follows in turn that, in iteration $k$ of the \pdr algorithm, all counterexamples traces (\autoref{def:counter}) of length $k$ have been eliminated, because otherwise \Fprop[1] would not hold.
Finally, whenever $F_i = F_{i+1}$ for some $i < k$, then
 \Fprop[3] implies that $F_i$ is an inductive invariant.
Because \pdr maintains the candidates $F_i$ as CNF formulas (the candidates are refined by blocking cubes, which is the same as conjoining clauses),
this termination check can be done syntactically~\cite{bradley2007checking} (without expensive SAT solver call).

Towards incrementing $k$, the algorithm initializes $F_{k+1}$ to $1 = S = \set{0,1}^X$ for each $k$.
To make candidate $F_{k+1}$ satisfy \Fprop[1], \pdr proceeds to remove states (block cubes) $s\in F_{k+1} \setminus P$ (initially $s\in \compl P$).
But removing these states might invalidate \Fprop[2] and \Fprop[3].
So the algorithm searches backwards to also refine previous candidates $F_i$.
Once a candidate $F_i$ (which overapproximates states reachable in $i$ steps) is strong enough to show that some $s\in F_{i+1}$ cannot be reached in one step, $F_{i+1}$ is refined by removing $s$ by constraining the candidate with the negation of cube $s$ (a clause).
We now explain this process in more detail.

\begin{definition}[Relative Inductivity] \label{def:rel-ind}
    A formula $F$ is \textit{inductive relative} to $G$ under a transition
    relation $\Delta$ if  $(F \cap G).\trans \subseteq F$  (or ${F} \land {G} \land \trans \land \neg
    {F}\primed = 0$).
\end{definition}

    \begin{definition} \label{def:frames}
        The \textit{sequence of candidate \iis}, or simply the
        \textit{candidates}, is defined as $F = \{\nary{F_0 = I, F_1, F_2,
        \ldots, F_k}\}$. It has the following properties \Fprop
        \cite{bradley2011sat}, which are maintained throughout the \pdr algorithm:
        \begin{align}
              & F_0  = I, \hfill
            \tag{\Fprop[0]} \\        	
            \forall 0 \le i \le k\colon & F_i \subseteq P, \hfill
            \tag{\Fprop[1]} \\
            \forall 0 \le i < k  \colon & F_i \subseteq F_{i+1},
            \tag{\Fprop[2]} \\
            \forall 0 \le i < k  \colon   & F_i.\trans \subseteq F_{i+1} \tag{\Fprop[3]}
        \end{align}
    \end{definition}\vspace{-1ex}

\begin{figure} [t]
    \centering
\vspace{-1em}
    \begin{tikzpicture}[initial text ={input: STS $(X, I, \Delta)$ and $P$ with $I\subseteq P$}]
        \tikzset{
            basic/.style={
              rectangle,
              align=center,
              text width=2.5cm,
              node distance=1.1cm and 1.7cm
              },
            narrow/.style={
              rectangle,
              align=center,
              text width=2cm,
              node distance=1cm and 1.7cm
              },
            vtag/.style={
              rectangle,
              align=center,
              text width=1.9cm,
              },
            vtagW/.style={
              rectangle,
              align=center,
              text width=2.9cm,
              },
            ltag/.style={
              rectangle,
              align=left,
              text width=1.9cm,
              },
            rtag/.style={
              rectangle,
              align=right,
              text width=1.3cm,
              },
            ltagW/.style={
              rectangle,
              align=left,
              text width=3.9cm,
              },
            over/.style={
              rectangle,
              align=center,
              text width=4.5cm,
              },
            forked/.style={to path={-| (\tikztotarget) \tikztonodes}}
        }
        
         \node (x) {input: STS $(X, I, \Delta)$, $P$ with $I\subseteq P$ and $F = \set{1,1, \dots}$};
    
        \node[draw, below= 1.cm of x,basic, text width=3cm, ] (0)  
          {\ttfunction{pdr-main}:\\ Maintain $F_0\ldots F_k$ according to \autoref{def:frames}};
        \node[draw, basic, below=.7cm  of 0] (5) {Done refining\\(full PDR would \concept{propagate} here)};
        \node[draw, basic, text width=3cm,  left =1cm  of 5] (1) {\ttfunction{block}: process queue and refine $F_i$ \\ (without \textit{generalize})};
        \node[draw, basic, text width=3cm, below=1.2cm of 1] (2) {Search deeper if possible};
        \node[draw, basic, line width=2pt, right = of 2] (6) {Counterexample};
        \node[draw, narrow, line width=2pt, right=.6cm of 5] (7) {Inductive invariant $F_i$};
        
        \draw[-latex] (x) edge node[left,text width=2.8cm,] {Set $k := 0,$ $F_0 := I$} (0);
        
        \draw[-latex] (0) -| node[above, text width=4.5cm, xshift=-1em] {\textit{Otherwise: no invariant\\ Initialize queue $O := \set{( \compl P, k)}$}} (1);
        \draw[-latex] (0) -| node[vtagW, above, xshift=0em] 
          { $\exists i\colon F_i = F_{i+1}$ \textit{($F_i$ is an inductive invariant)}} (7);
        
        \draw[-latex] (1) edge[bend right] node[left,text width=2.5cm] 
        	{$\exists (s, i) \in O, t \in F_i$ \\ with $(t,s) \in \Delta$} (2);

        \draw[-latex] (1) edge[loop left] node[above = .6cm, xshift=1.5cm, text width=4.cm] 
        	{$\exists (s, i) \in O$ with $s \notin F_i.\Delta$\\
        	then set  $O := O \setminus \set{ (s, i)},$\\
        	$\forall j=1..i\text+1 \colon F_{j} := F_{j} \setminus \set s$} (2);
        	
        \draw[-latex] (1) edge node[vtag, above] {$O \text= \emptyset$} (5);
        
        \draw[-latex] (2) edge[bend right] node[ltagW, right] 
          {if $i> 0 $ then \\ set $O := O \cup (t, i-1)$} (1);
        \draw[-latex] (2) edge node[vtag, below] {if $i=0$} (6);
        
        \draw[-latex] (5) edge node[ltag, right] {\textit{increment k}} (0);
    \end{tikzpicture}
    
    \caption{A simplified \pdr algorithm}
    \label{fig:overview}
\vspace{-1em}
\end{figure}
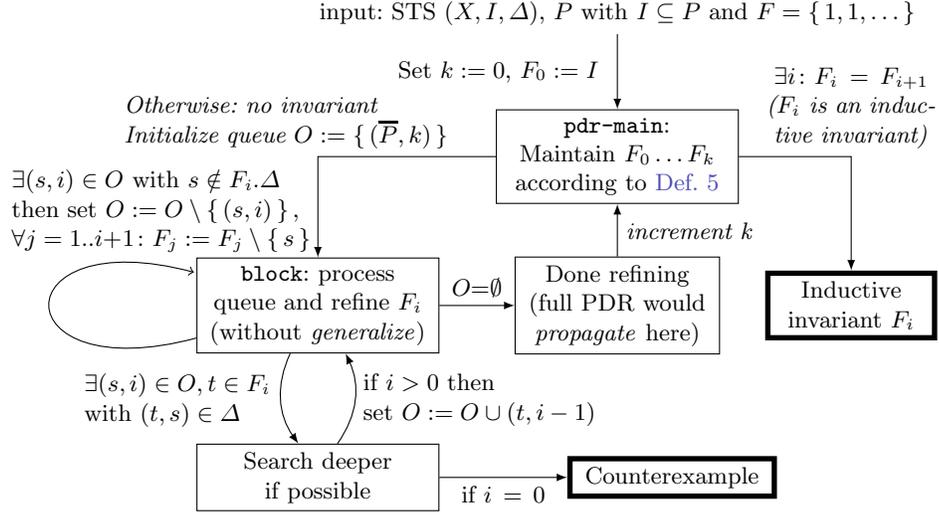

\autoref{fig:overview} gives the simplified \pdr algorithm.
\pdr takes as inputs: an STS according to \autoref{def:ts}, a property $P$ and a sequence of candidate inductive invariants $F$, all initialized to 1. It then produces either an \ii, which proves $P$ to be an invariant of the system,
or a \textit{counterexample trace}, i.e., a path that shows a violation of $P$ (see \autoref{def:counter}).
To do this, \pdr
iteratively extends the sequence of candidate \ii{s} \frames[k] in a major loop (\texttt{pdr-main}).
To maintain all invariants in \autoref{def:frames}, the candidates are refined by a minor loop (\texttt{block}) within each major loop iteration. 
This loop uses a queue $O$ of \concept{proof obligations} $(s, i) \in S \times \mathbb N$;
obligations to show that a state $s$ is not reachable in $i+1$ steps.
If there is no $t\in F_i$ that can transit to $s$ ($\Delta(t,s) = 1$), then
$(s,i)$ is removed from $O$ and $s$ is blocked in all candidates
$F_1, F_2, \dots F_{i+1}$, preserving \Fprop[3] (because the algorithm never
adds an obligation $(u,j)$ with $u \in I$ as this would constitute one end of a counterexample trace).
The queue $O$ is initialized with $(\compl P, k)$, because the algorithm wishes to refine $F_{k+1}$ until it is a subset of $P$ so that \Fprop[1] is satisfied when $k$ is increased to $k+1$.\footnote{Without loss of generality, we may assume that $\compl P$ is a single state (i.e., a sink state).
}
Once the minor loop completes the search by refining candidates, the major loop continues by incrementing $k$.

\begin{definition}[Counterexample trace] \label{def:counter}
    A \textit{counterexample trace} for an STS $(X,I,\Delta)$ and a property $P$ is a path $\pi_0, \pi_1, \dots, \pi_m \in S$ 
with   $\pi_0 \in I$ and $\pi_m \in \compl P$.
\end{definition}

We now extend \pdr with an internal state for restarting the algorithm.
    The goal of the internal state is to let \pdr suspend the search upon encountering a counterexample or an inductive invariant, to restart it later on a constrained or relaxed instance of the STS.   
\pdr uses two main data structures during its execution the sequence of
    candidate \iis~$F$ and a priority queue~$O$ to track outstanding
    proof-obligations.
\autoref{def:pdr-state} gives a \concept{valid \pdr state} that also records invariants maintained by the proof obligation queue $O$.

    \begin{definition}[Valid \pdr state] \label{def:pdr-state}
        Given an STS $M = (X, I, \Delta)$ and a property $P\subseteq \set{0,1}^X$, a \textit{valid \pdr state} is a tuple $(M, P, k, F, O)$ that satisfies:
        \begin{itemize}
            \item $F = F_0, F_1, \dots, F_k$ with $0 \leq k < 2^{\sizeof{X}}$ is a sequence of candidate invariants that adheres to the properties \Fprop from \autoref{def:frames}, and
            \item $O \subseteq  {S\times \mathbb N}$ is a queue (set) of proof obligations adhering to the properties:\vspace{-1ex}
                        \begin{align}
            \forall \pair{s}{i} \in O\colon 
                & 0 < i \le k \tag{\Oprop[1]} \\
             \forall \pair{s}{i} \in O\colon    &  (s,i ) = (\compl P, k) ~ \lor ~ 
             	\exists (t, i+1) \in O \colon \Delta(s,t) = 1
              \tag{\Oprop[2]}\\
             \forall \pair{s}{i} \in O\colon    & s\notin F_i
                  \tag{\Oprop[3]}  
        \end{align}
        \end{itemize}
    \end{definition}

    The properties \Oprop follow from the fact that the minor loop (\texttt{block}) 
    basically performs a backwards search starting from $\compl P$ states.
	The property \Oprop[2] in particular requires that all proof obligations lead to a $\compl{P}$ state (via other proof obligations).
    Only \Oprop[3] is non-trivial in this respect: It follows from the fact that in each major iteration $k$ (\texttt{pdr-main}) the nonexistence of  counterexample traces of length $k$ has been proved, as noted above. Now if $s$ would be in $F_{i}$, a counterexample trace of length $k$ would exist, thus contradicting the \Fprop invariants of the algorithm.

    \begin{algo}[b]
    \caption{Propagation at the level of formulas and SAT  queries} 
    \DontPrintSemicolon
\SetKwFunction{Fpropa}{propagate}
\SetKwFunction{Fhif}{highest\_inductive\_frame}
\SetKwFunction{Fgen}{generalize}
\SetKwFunction{Fremove}{remove\_cube}
\SetKwFunction{Fmic}{generalize}
\SetKwFunction{Fdown}{down}
\SetKwFunction{Fblock}{block}
\SetKwFunction{Ftop}{top}
\SetKwFunction{Fmax}{max}
\SetKwFunction{Fstrenghten}{strengthen}
\SetKwFunction{Fpdr}{refine}
\SetKwFunction{Fstrategy}{pdr\_strategy}
\SetKwFunction{Fipdrmain}{pdr-main}
\SetKwFunction{Fipdrconstrain}{ipdr\_constrain}
\SetKwFunction{Fipdrrelax}{ipdr-relax}
\SetKwFunction{Fipdrconstrainloop}{constrain}
\SetKwFunction{Fipdrrelaxloop}{relax}
\SetKwFunction{Fsubsume}{remove\_subsumed}

\SetKwFunction{Fpeter}{peterson}
\SetKwFunction{Fgraph}{graph-search}

\SetKw{KwST}{\ s.t.\ }
\SetKwFor{Loop}{loop}{}{end}

\SetKwInOut{In}{In}
\SetKwInOut{Out}{Out}
\SetKwInOut{Pre}{Pre}
\SetKwInOut{Post}{Post}
\SetKwInOut{Vars}{Vars}

	\label{alg:pdr:propagate}

	\In{
			A sequence $\nary{F_0, F_1, \ldots, F_{k}}$ (in CNF form $\form F_i$) and STS $M = (X, I, \Delta)$.	
	}\vspace{-1ex}

\func{\Void}{\Fpropa{$F,\   \ts$}} {\Comment*{CNF $\form F_i$ (set of clauses) represents $F_i$: \vspace{-1.2em}}
    \For{$i \gets 1$ \KwTo $k-1$}{  \Comment*{Find the last $\form F_i$ containing $\form C$ \vspace{-1.2em}}
        \ForAll{$\form C \in \form F_i \setminus \form F_{i+1}$ }{ 
			\If{$\SAT(\form F_i \land \Delta \land \neg \form C') = 0$} { \label{l:prop:push-query}
  				$\form F_{i+1} \gets \form F_{i+1} \cup \set{\form C}$\;  \label{l:prop:remove}
  			}\vspace{-1ex}
  		}\vspace{-1ex}
    }\vspace{-.5ex}\Comment*{Modified indirectly through its formula representation \form F \vspace{-1.2em}}
    \Return{F}
}\vspace{-1em}
\end{algo}
~
\vspace{-1em}
\vspace{-1em}
\vspace{-1em}

The full version of \pdr adds generalization of states and propagation of clauses in candidates $F_i$, which we briefly explain  here, as \ipdr uses propagation as well.
In the above, the refinement of a candidate $F_i$ is done by blocking (removing) a state
$F_i := F_i \setminus \set s$ (in set notation). In reality, \pdr maintains $F_i$ as a CNF formula~$\form F_i$; a conjunction of clauses $\form C$ which represent the removed states~$\neg \form C$.
However, before blocking, states are first generalized by dropping literals from~$\form C$.
The generalization $g_s$ of state $s$ can greatly strengthen a candidate $F_i$
by blocking (removing) many states at once (since  $g_s$ is a subcube of $s$, we have $s \implies g_s$) that can all be proven unreachable in $i$ steps (while taking care not to remove initial states!).
So each (blocking) clause in $\form F_i$ is a negated cube $g_s$, a subcube of $s$.

Generalization enables propagation (see \autoref{alg:pdr:propagate}), as now other blocking clauses in $\form F_i$ may be used to strengthen later candidates $\form F_{i+1}$. Propagation pushes blocking clauses forward and is done during the minor search (\texttt{blocking}) and after it completes (see note in \autoref{fig:overview}).
Both generalization and propagation have been shown to preserve the \Fprop invariants~\cite{bradley2007checking}. They also do not affect the proof obligation queue. Therefore, \autoref{def:pdr-state} holds in the full \pdr algorithm.

We can now extend \pdr with the internal state $Y = ((X, I, \Delta), P, k, F, O)$ from \autoref{def:pdr-state}, by modifying the inputs and initializations on the initial edge in \autoref{fig:overview}.
The inputs and variables are now all set to those provided by a valid internal state (obtained from a previous run).
In the first \ttfunction{pdr-main} loop iteration, $O$ should also not be reset to
$\set{(\compl P, k)}$ but instead taken from $Y$.
When the algorithm terminates with a counterexample or an invariant, it should also return the current internal \pdr state.
    Note that the valid \pdr state does not record the line at which the \pdr algorithm currently is. Consequently, we will be able to restart it with a modified but valid \pdr state to obtain a different result (e.g., to avoid a counterexample in the next run by further constraining the system).

\vspace{-1ex}
\subsection{Incremental Property Directed Reachability (\ipdr)}
\vspace{-1ex}
\label{ss:ipdr}

 \ipdr takes a sequence of constraining (or relaxing) STS instances $M_1, \dots, M_z$ such that $M_{i+1} \sqsubseteq M_{i}$ (or $M_{i} \sqsubseteq M_{i+1}$), solving them one by one with the extended \pdr algorithm (\ttfunction{pdr-main}), while passing the internal state along to~speed up subsequent \pdr runs. 
Here we define relaxing (and constraining) \ipdr as a loop around the \pdr algorithm (extended with internal state).
In relaxing \ipdr, the outer loop terminates when \pdr finds a  counterexample trace. But when \pdr finds an inductive invariant for the current instance $M_i$ (wrt property $P$), \ipdr relaxes the instance to some $M\relaxed = M_{i+1} \sqsupseteq M$ and calls \pdr iteratively.
In constraining \ipdr, the algorithm instead terminates when an inductive invariant is found and iterates on a constrained version of the STS when a counterexample is found.
In both cases, we show how to modify the internal \pdr state such that is a valid internal \pdr state for the constrained or relaxed system.

	\paragraph{Constraining \ipdr Algorithm}
		\autoref{ipdr:constrain} shows constraining \ipdr.
		Initially, \ttfunction{pdr-main} is called on the system $M_1$ wrapped in a valid \pdr state (\autoref{l:pdrcall1}). Effectively, \ttfunction{pdr-main} is initialized like in \autoref{fig:overview}.
		Then the algorithm considers the constrained systems $M_2, \dots, M_z$ iteratively.
		If the previous \ttfunction{pdr-main} returned an inductive invariant, then the algorithm stops at \autoref{l:ce}, as further constraining the system is not necessary.\footnote{This represents for instance the scenario when we find the minimum number of pebbles to successfully pebble a circuit by approximating the pebble count from above; reducing pebbles in each run until goal of pebbling the circuit is no longer possible.}
		Otherwise, a more constrained system $M_i \sqsubset M$ or $M_{c+1} \sqsubset M$ is considered next
		(we may skip instances $M_{i}..M_c$ when a counterexample  is found that is valid in $M_c$ for $c \ge i$, as \autoref{l:forward} does).
		However, we first update the \pdr state for the constrained STS $M\constrained = M_i$ using the \ttfunction{constrain} operation.
		It resets $O$ at \autoref{l:oprop2} to satisfy \Oprop in the constrained system $M\constrained$ (repairing $O$ would require at least as many operations as simply restarting the search).
		It also propagates blocking clauses between candidates $F_1, \dots, F_k$ as constraining can potentially block them.
		Finally, \ttfunction{pdr-main} is called again for the updated \pdr state and the \ipdr continues to the next iteration (\autoref{l:pdrcall2}).

		Constraining a system does not add behavior, therefore all the \Fprop invariants remain intact (intuitively:
			\emph{for the constrained system}, the candidates $F_i$ still are valid over-approximations of the reachable states in $i$ steps and they disprove counterexamples of length $i$).
			The propagation does not change this, as it merely strengthens frames while preserving~\Fprop.
		Because the queue $O$ is emptied, $\Oprop$ holds vacuously.
We conclude that the \ttfunction{constrain} function indeed yields a new valid \pdr state according to \autoref{def:pdr-state}. Consequently, the iterative call to \ttfunction{pdr-main} at \autoref{l:pdrcall2} of \autoref{ipdr:constrain} can proceed incrementally checking the constrained system.

	\paragraph{Relaxing \ipdr Algorithm} \label{ipdr:relax:alg}
		\autoref{alg:ipdr:relax-prep} shows relaxing \ipdr. Like in the constraining version,
		initially, \ttfunction{pdr-main} is called on the system $M_1$ wrapped in a valid \pdr state (\autoref{l:relax:pdrcall1}). Effectively, \ttfunction{pdr-main} is initialized like in \autoref{fig:overview}.
		Then the algorithm considers the constrained systems $M_2, \dots, M_z$ iteratively.
		Now, if the previous \ttfunction{pdr-main} returned a counterexample trace, then the algorithm stops at \autoref{l:relax:ce}, as further relaxing the system is not necessary.\footnote{This represents for instance the scenario when we find a bug after increasing the number of interleavings in a parallel program.}
		Otherwise, the relaxed system $M_i \sqsubset M$ is considered next using the valid \pdr state from the previous run.
		However, we first update the \pdr state for the constrained STS $M\constrained = M_i$ using the \ttfunction{relax} operation. 
		Here, it constructs a new set of candidate inductive invariants $F$.
		Finally, \ttfunction{pdr-main} is called again for the updated \pdr state and the \ipdr continues to the next iteration (\autoref{l:relax:pdrcall2}).

\begin{algo}[p]

\newcommand{\alg}{alg:ipdr:constrain-prep}
\caption{Constraining \ipdr (\cipdr)}
\label{ipdr:constrain}

\In{\pdrstate with $F = \set{F_0, \ldots, F_{k}}$ satisfying \autoref{def:pdr-state} and
     $ M\constrained \sqsubset  M$ 
}
\Out{A valid \pdr \textit{state} for STS $M\constrained$ according to  \autoref{def:pdr-state} }
\vspace{-1.5ex}

\func{\Trace}{\Fipdrconstrainloop{$\pdrstate,\ M\constrained$}}{

    $F\constrained \gets \nset{I\constrained, F_1, F_2, \ldots, F_{k}}$ \;
    
	  $F\constrained \gets $~\Fpropa{$F\constrained, M\constrained$} \Comment*{See \autoref{alg:pdr:propagate} \vspace{-0em}}
	   \label{\alg:prop}

	
    \Return{$ \tuple{M\constrained, P,  k,  F\constrained, O := \emptyset}$}
    \Comment*{Repair $\Oprop[2]$ from \autoref{def:pdr-state} by setting $O := \emptyset$}\label{l:oprop2}
}


\vspace{-1ex}
\In{$ M_1 \sqsubset M_2 \sqsubset \dots \sqsubset M_z$ with $M_i = (X, I_i,\ \trans{}_i)$ and
	$P\subseteq \bool^X$
}
\Out{
A counterexample trace or inductive invariant
}\vspace{-1.5ex}

\func{\Trace}{\Fipdrconstrain{$M_1, M_2, \dots M_z$, P }}{
	$F \gets \set{F_0 := I_1, F_1 := 1}$\;
	\pdrstate, \textit{result} $\gets$ \Fipdrmain{$(M_1, P, k := 0, F, O := \emptyset)$ 
												}\;\label{l:pdrcall1}
	\For{$M_i \in \set{M_2, \dots M_z}$}{
		\If{result is an inductive invariant}{\vspace{-.5ex}
			\Return{result}\vspace{-.5ex}								\label{l:ce}
		}
		\If{result is a trace valid in $M_c$ for $c\ge i$}{\vspace{-.5ex}
			\textbf{forward loop} to $M_i := M_{c+1}$\vspace{-.5ex}			\label{l:forward}
		}
		\pdrstate\phantom{, result}$\gets$ \Fipdrconstrainloop{\pdrstate, $M_i$}\;
		\pdrstate, result $\gets$ \Fipdrmain{\pdrstate}\label{l:pdrcall2}
	}\vspace{-1ex}
	\Return{result}
}
\vspace{-1em}
\end{algo}
%
%
%
%
%
\begin{algo}[p]

\caption{Relaxing \ipdr (\ripdr)}
\label{alg:ipdr:relax-prep}

\In{\pdrstate with 
				 $F = \set{F_0, .., F_{k}}$ satisfying \autoref{def:pdr-state} and ~~~~~~~~~~~~~~~~~~~~
     $M\relaxed \sqsupset  M$ with $M\relaxed = (X, I\relaxed,\ \trans\relaxed)$.
}

\Out{A valid \pdr \textit{state} for STS $M\relaxed$ according to \autoref{def:pdr-state} }
\vspace{-1.5ex}

\func{\Trace}{\Fipdrrelaxloop{$\pdrstate, M\relaxed$}}{

     $F\relaxed \gets \set{F_0 := I\relaxed,  F_1  :=  1, \dots , F_k  := 1}$ 
    \Comment*{The tautology 1  is $\emptyset$ in CNF}
        \label{alg:relax:reprop}
	
	 \Comment*{As \autoref{alg:pdr:propagate}; try copy clauses from $\form F_i$ to $\form F_i\relaxed$\vspace{-1.3em}}
    \For{$i \gets 1$ \KwTo $k-1$}{  \label{l:relax:prop}
        \ForAll{$\form C \in \form F_i$ }{  \vspace{-1em}\Comment*{Access the candidates $F$ in CNF form \form F}
    \Comment*{Does $\neg\form C$ not 
    block new initial states? \vspace{-1.3em}}
	    	\If{$\SAT(\form I\relaxed \land \neg \form C) = 0$} { 
			\lIf{$\SAT(\form F_i\relaxed \land \Delta\relaxed \land \neg \form C') = 0$} {  			
				$\form F_{i+1}\relaxed \gets \form F_{i+1}\relaxed \cup \set{\form C}$
  			}
  			}\vspace{-1ex}
  		}\vspace{-1ex}
    }
        
%
%
        \Return{$ 
            \tuple{M\relaxed, P, k := 0,  F\relaxed, O}$} \label{l:resetk}
}
\vspace{-1.ex}

%
%
%
%


\In{$ M_1 \sqsupset M_2 \sqsupset \dots \sqsupset M_z$ with $M_i = (X, I_i,\ \trans{}_i)$ and $P\subseteq \bool^X$
}
\Out{A counterexample trace or inductive invariant}\vspace{-1.5ex}
\func{\Trace}{\Fipdrrelax{$M_1, M_2, \dots M_z$, P }}{
	$F \gets \set{F_0 := I_1, F_1 := 1}$\;
	\pdrstate, \textit{result} $\gets$ \Fipdrmain{$(M_1, P, k:=0, F, O := \emptyset)$
											}\;\label{l:relax:pdrcall1}
	\For{$M_i \in \set{M_2, \dots M_z}$}{
		\If{result is a counterexample trace}{\vspace{-.5ex}
			\Return{result}\label{l:relax:ce}\vspace{-.5ex}
		}
        \If{$\exists s \in I_i\relaxed \cap \compl P$}{\label{l:relax:check}\vspace{-.5ex}
            \Return{\textit{trace} $(s)$}\vspace{-.5ex}
        }
		\pdrstate\phantom{, result}$\gets$ \Fipdrrelaxloop{\pdrstate, $M_i$}\;
		\pdrstate, result $\gets$ \Fipdrmain{\pdrstate}\;\label{l:relax:pdrcall2}
	}\vspace{-1ex}
	\Return{result}
}\vspace{-1em}
\end{algo}


		After an instance terminates in iteration $k$, the \ttfunction{pdr-relax} function checks whether the \Fprop  properties hold for $i=0$ at \autoref{l:relax:check}. If not, \ipdr found a short counterexample because of newly introduced (and erroneous) initial states in the relaxed instance $M\relaxed = M_i$ and terminates.
As relaxing introduces new transitions, the candidates $F_i$ may no longer be strong enough to prove unreachability of states $\neg \form C$ for each $\form C \in \form F_{i+1}$. 
Therefore, the new \pdr-state created by the \ttfunction{relax} function will have to begin from $k=0$ again in order to strengthen the frames enough to prove unreachability of $\overline P$ in multiple steps.
Nonetheless, \ttfunction{relax} attempts to preserve as much as possible of the old candidate sequence 
 in a new sequence  $F\relaxed$, so that once the \pdr run on the relaxed system increases $k$, it potentially no longer starts with a frame $F_{k+1} = 1$ (or equivalently $\form F_i= \emptyset$),
			but with a subset of the blocking clauses from the previous \pdr run.
			This can be done through a mechanism similar to the propagation phase
			from \autoref{alg:pdr:propagate}. From \autoref{l:relax:prop} in \ttfunction{relax}, blocking clauses \form C in the original candidates $\form F_i$ (the CNF form of $F_i$) are inspected and copied to $\form F_i\relaxed$ if two conditions hold:
			1) $\neg \form C$ does not block a new initial state in $I\relaxed$,
			and 2) the candidate $\form F_{i-1}\relaxed$ is strong enough to prove unreachability of $\neg \form C$ in the relaxed STS (i.e., under transition relation~$\Delta\relaxed$).

			Because $k$ is set to $0$, \Fprop trivially holds, but the pre-initialized frames are also sound for $\Delta\relaxed$.
			Because relaxing \ipdr only considers a \pdr-state tuple when \ttfunction{pdr-main} terminates with an invariant, all outstanding obligations have been
			eliminated ($O = \emptyset$) before returning the inductive invariant (see \autoref{fig:overview}). This vacuously satisfies \Oprop.
We conclude that the \ttfunction{relax} function indeed yields a new valid \pdr state satisfying  \autoref{def:pdr-state}. Consequently, the iterative call to \ttfunction{pdr-main} at \autoref{l:relax:pdrcall2} of \autoref{alg:ipdr:relax-prep} can proceed incrementally checking the relaxed system.

\paragraph{Binary Search with Relaxing and Constraining \ipdr} 
Assuming the value of the target optimization parameter equals $p$, e.g., the minimal number of pebbles required to pebble a circuit, relaxing \ipdr needs $p$ \pdr calls to find it.
Assuming a sound upper bound $b$ on $p$, e.g., the number of gates in the circuit, constraining \ipdr takes (at least) $b - p$ calls.
By combining the \ttfunction{pdr-relax} and \ttfunction{pdr-constrain} functions, a binary search algorithm (\bipdr) takes only $\log(b)$ \pdr calls, or $O(\log(p))$ in practice, since often $b = c\cdot p$~\cite{meuli2019reversible,quist}. We omit the details of \bipdr here,
as the algorithm boils down to a standard binary search that combines the \texttt{constrain} and \texttt{relax} routines of \autoref{ipdr:constrain} and \autoref{alg:ipdr:relax-prep}.

\section{Related Work}
\label{sec:related}

The IC3 algorithm was introduced in~\cite{bradley2011sat,bradley2007checking} and later extended into Property Directed Reachability (\pdr) in~\cite{efficient-pdr,Welp2013,Hoder2012}.
To deal with infinite-state systems, \pdr versions~\cite{birgmeier2014counterexample,hoder2012generalized,cimatti2012software,vvt} exists that use SMT~\cite{barrett2009handbook} and abstraction-refinement~\cite{clarke2000counterexample,beyer2013explicit}.
The works \cite{goel2021symmetry} and \cite{gunther2017dynamic} extend \pdr with transaction and symmetry reduction.

Well-structured transitions systems~\cite{finkel2001well} provide another formalization of relaxed and constrained systems (\autoref{def:relaxconstrain}) that has been used to verify infinite-state systems like priced~\cite{larsen2001cheap} and timed automata~\cite{alur1999timed,timedltl}.

Context-bounded analysis~\cite{cba,qadeer2005context,cordeiro2011verifying} in concurrent programming deals with the study of programs by adding restrictions on the context switches of threads.
Incrementally increasing parallel interleavings has also been exploited for model checking in~\cite{Grumberg:2005:PUM:1040305.1040316,context-switch}.

Pebble games are \PSPACE-complete and have a long history in computer science~\cite{chan2013pebble,lingas1978pspace}.
Reversible pebble game optimization was studied in~\cite{meuli2019reversible,quist}.

\section{Implementation and Experimental Evaluation}
\label{sec:experiments}
\label{sec:implementation}

\paragraph{\ipdr Implementation}
We implemented \ipdr in relaxing, constraining and binary search form.
The open source implementation in C++ is available at GitHub.\footnote{\url{https://github.com/Majeux/pebbling-pdr}}
It uses the SAT solver Z3~\cite{z3} and fully exploits its incremental solving capabilities for \pdr (internally) and \ipdr (in between incremental runs).  
It contains the following optimizations that have been discussed before for IC3 and \pdr. (None of which interfere with the discussed modifications for \ipdr.)
\begin{itemize}
	\item The delta encoding of \cite{efficient-pdr} avoids duplicating blocked clauses by only storing blocked clauses for the highest frame where it occurs.
	\item Subsumbtion checks \cite{bradley2011sat,efficient-pdr} avoid storing redundant weaker blocked clauses.
	\item Generalization~\cite{bradley2011sat} with the later extension with the \texttt{down} algorithm~\cite{ic3-ctg} finds stronger clauses to block. 
	This methods brings along some additional parameters \texttt{ctgs} and  \texttt{max-ctgs}. After some preliminary testing, these parameters were set to 1 and 5 respectively; in line with the findings Bradley \cite{ic3-ctg}.
	\item Before handling a proof-obligation in the minor \pdr loop, a
		subsumption check  can quickly
		detect if newly added clauses already
		block it \cite{efficient-pdr}.
	\item We also preempt future obligations by re-queueing a proof obligation $(s,i)$ as $(s,i+1)$, since it will have to be proven in later iterations anyway~\cite{efficient-pdr}.
\end{itemize}

\paragraph{Benchmarks}
As discussed in the introduction, we choose to apply \ipdr to the optimization problem of reversible circuit pebbling. 
We encoded the transition relations as described in~\cite{meuli2019reversible}.
For a number of pebbles $p$ between 1 and $g$ (the number of gates in the circuit), we encode a separate system $M_p$. It is easy to see that adding more pebbles relaxes the problem, i.e., $M_i \sqsubset M_{i+1}$.
We took circuits from the
	Reversible Logic Synthesis Benchmarks website \cite{rls-benchmarks},
	which lists several ``families" of circuits of increasing size
	and complexity.
	We selected those circuits that could be completed within half an hour by all benchmarked algorithms.

For experimenting with increasing the number of interleavings~\cite{Grumberg:2005:PUM:1040305.1040316,context-switch,clarke-grumberg-minea-peled}, we encoded the Peterson mutual exclusion protocol~\cite{peterson}.
We added a scheduler~\cite{context-switch} to the encoding to bound the number of interleavings~to $\ell$, starting from zero.

\paragraph{Experimental Setup}
All experiments were run on a computer with
16 GB of 2400 MHz DDR4 memory and an i7 6700T CPU.

All benchmarks were repeated ten times using a different random seed for the initialization of the underlying Z3 \sat-solver.
We compare \ipdr with our own naive \pdr implementation that does not reuse information between incremental runs and with the the \pdr implementation of  Z3's \cite{z3}: \spacer \cite{spacer}, for which we re-encoded the systems in Horn clauses.

We first investigate  whether \ipdr can speed up parallel program verification by gradually increasing the instance hardness through controlling the number of allowed interleavings for the Peterson protocol.
Second, we investigate whether \ipdr can solve pebble optimization faster, by either: a) approaching the problem from below by relaxing the system (adding pebbles until a `counterexample' is found that representing a successful pebbling), b) approaching the problem from above by constraining the system with fewer pebbles starting from an upper bound 
	(e.g., when the number of pebbles equals the number of gates in the circuit, so the circuit can trivially be pebbled by covering all gates), or c) combining both approaches in binary search.
Third, we aim to gather statistics to study the amount of information that \ipdr can   reuse between runs.

\paragraph{Results for Peterson}
\autoref{fig:peter:relax:all} shows the runtimes for
the Peterson protocol with 2, 3 and 4 processes with a timeout of four hours.
The number of context switches that was
feasible to run: \ttfunction{Peterson2} was verified up to a bound of 10
switches, \ttfunction{Peterson3} to a bound of 4 and \ttfunction{Peterson4}
to 3.
With our Horn clause encoding, \spacer is not competitive.
With the most incremental steps, \ttfunction{Peterson2} achieved a speedup of almost a factor four over naive \pdr. \ttfunction{Peterson3} and \ttfunction{Peterson4} both
showed a similar improvement of around a factor 1,7.

\begin{figure}[p!]
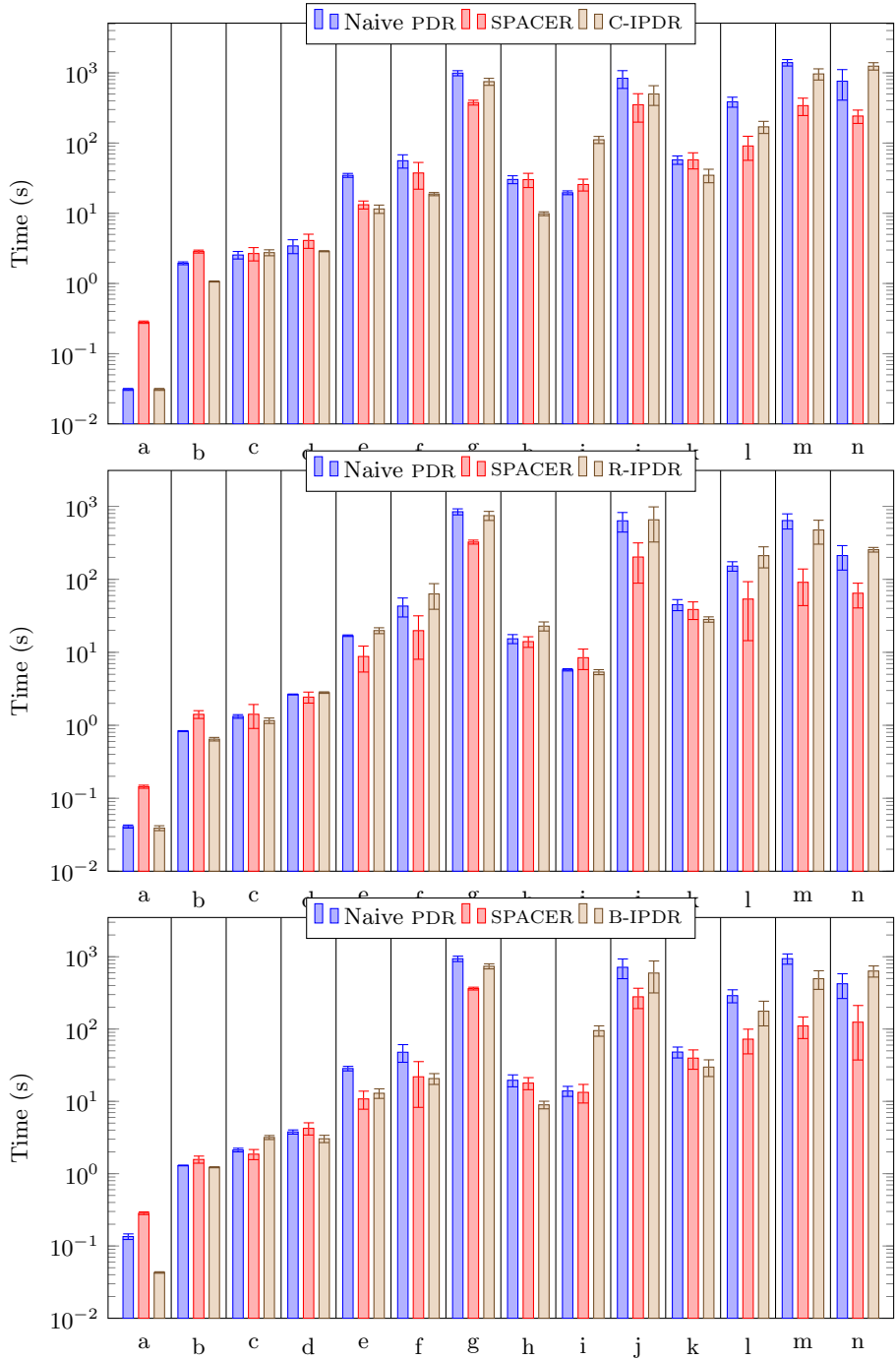

 \centering

\begin{vsplot}{Naive \pdr, \spacer, \cipdr}{
    a,
    b,
    c,
    d,
    e,
    f,
    g,
    h,
    i,
    j,
    k,
    l,
    m,
    n
}

    \addplot+[
        error bars/.cd,
        y dir=both,
	    y explicit
    ] coordinates {
		(a, 0.031) +- (0, 0.001)	
		(b, 1.942) +- (0, 0.092)	
		(c, 2.545) +- (0, 0.312)	
		(d, 3.428) +- (0, 0.768)	
		(e, 34.711) +- (0, 2.240)	
		(f, 55.970) +- (0, 12.034)	
		(g, 989.706) +- (0, 82.501)	
		(h, 30.292) +- (0, 3.897)	
		(i, 19.618) +- (0, 1.233)	
		(j, 836.655) +- (0, 237.205)	
		(k, 57.610) +- (0, 7.894)	
		(l, 386.719) +- (0, 62.920)	
		(m, 1396.920) +- (0, 146.996)	
		(n, 760.006) +- (0, 351.693)	
    };

     \addplot+[
         error bars/.cd,
         y dir=both,
         y explicit
     ] coordinates {
		(a, 0.281) +- (0, 0.011)	
		(b, 2.850) +- (0, 0.139)	
		(c, 2.670) +- (0, 0.574)	
		(d, 4.098) +- (0, 0.934)	
		(e, 13.225) +- (0, 1.721)	
		(f, 37.466) +- (0, 15.443)	
		(g, 377.234) +- (0, 29.790)	
		(h, 30.153) +- (0, 6.907)	
		(i, 25.678) +- (0, 4.931)	
		(j, 351.181) +- (0, 152.675)	
		(k, 57.699) +- (0, 14.907)	
		(l, 90.823) +- (0, 34.076)	
		(m, 341.008) +- (0, 94.748)	
		(n, 242.379) +- (0, 52.130)	
	};

	\addplot+[
		error bars/.cd,
		y dir=both,
		y explicit
	] coordinates {
		(a, 0.031) +- (0, 0.001)	
		(b, 1.067) +- (0, 0.018)	
		(c, 2.751) +- (0, 0.268)	
		(d, 2.878) +- (0, 0.048)	
		(e, 11.513) +- (0, 1.563)	
		(f, 18.761) +- (0, 0.973)	
		(g, 748.969) +- (0, 85.701)	
		(h, 9.865) +- (0, 0.620)	
		(i, 111.350) +- (0, 12.897)	
		(j, 499.460) +- (0, 156.197)	
		(k, 34.781) +- (0, 7.563)	
		(l, 170.244) +- (0, 33.414)	
		(m, 964.889) +- (0, 174.493)	
		(n, 1243.961) +- (0, 148.944)	
	};
\end{vsplot}

\begin{vsplot}{Naive \pdr, \spacer, \ripdr}{
    a,
	b,
	c,
	d,
	e,
	f,
	g,
	h,
	i,
	j,
	k,
	l,
	m,
	n
}
    \addplot+[
        error bars/.cd,
        y dir=both,
        y explicit,
    ] coordinates {
    	(a, 0.041) +- (0, 0.002)	
    	(b, 0.832) +- (0, 0.013)	
    	(c, 1.323) +- (0, 0.076)	
    	(d, 2.647) +- (0, 0.036)	
    	(e, 16.901) +- (0, 0.419)	
    	(f, 43.018) +- (0, 12.519)	
    	(g, 845.274) +- (0, 80.059)	
    	(h, 15.325) +- (0, 2.167)	
    	(i, 5.755) +- (0, 0.204)	
    	(j, 634.672) +- (0, 189.779)	
    	(k, 45.037) +- (0, 7.776)	
    	(l, 151.867) +- (0, 22.490)	
    	(m, 639.886) +- (0, 150.129)	
    	(n, 211.999) +- (0, 78.052)	
    };
    \addplot+[
        error bars/.cd,
        y dir=both,
        y explicit,
    ] coordinates {
    	(a, 0.144) +- (0, 0.007)	
    	(b, 1.413) +- (0, 0.173)	
    	(c, 1.416) +- (0, 0.511)	
    	(d, 2.433) +- (0, 0.420)	
    	(e, 8.773) +- (0, 3.405)	
    	(f, 19.833) +- (0, 11.799)	
    	(g, 324.640) +- (0, 20.350)	
    	(h, 14.018) +- (0, 2.331)	
    	(i, 8.441) +- (0, 2.639)	
    	(j, 202.320) +- (0, 113.627)	
    	(k, 38.639) +- (0, 10.474)	
    	(l, 53.712) +- (0, 39.254)	
    	(m, 91.152) +- (0, 47.485)	
    	(n, 64.588) +- (0, 24.033)	
	};
    \addplot+[
		error bars/.cd,
		y dir=both,
		y explicit
	] coordinates {
		(a, 0.039) +- (0, 0.003)	
		(b, 0.646) +- (0, 0.036)	
		(c, 1.159) +- (0, 0.099)	
		(d, 2.806) +- (0, 0.069)	
		(e, 19.870) +- (0, 1.839)	
		(f, 63.133) +- (0, 24.225)	
		(g, 747.978) +- (0, 106.465)	
		(h, 22.853) +- (0, 3.234)	
		(i, 5.366) +- (0, 0.423)	
		(j, 654.798) +- (0, 329.855)	
		(k, 28.258) +- (0, 2.370)	
		(l, 211.390) +- (0, 67.712)	
		(m, 474.993) +- (0, 171.428)	
		(n, 255.076) +- (0, 18.485)	
	};
\end{vsplot}

\begin{vsplot}{Naive \pdr, \spacer, \bipdr}{
    a,
    b,
    c,
    d,
    e,
    f,
    g,
    h,
    i,
    j,
    k,
    l,
    m,
    n
}

    \addplot+[
        error bars/.cd,
        y dir=both,
        y explicit,
    ] coordinates {
        (a, 0.135) +- (0, 0.012)	
        (b, 1.297) +- (0, 0.014)	
        (c, 2.124) +- (0, 0.123)	
        (d, 3.776) +- (0, 0.252)	
        (e, 28.433) +- (0, 2.048)	
        (f, 47.795) +- (0, 13.111)	
        (g, 938.345) +- (0, 80.643)	
        (h, 19.572) +- (0, 3.680)	
        (i, 13.916) +- (0, 2.201)	
        (j, 714.732) +- (0, 217.204)	
        (k, 48.104) +- (0, 8.263)	
        (l, 289.942) +- (0, 59.253)	
        (m, 941.210) +- (0, 152.290)	
        (n, 423.146) +- (0, 158.605)	
    };
    \addplot+[
        error bars/.cd,
        y dir=both,
        y explicit,
    ] coordinates {
        (a, 0.284) +- (0, 0.012)	
        (b, 1.573) +- (0, 0.177)	
        (c, 1.861) +- (0, 0.297)	
        (d, 4.239) +- (0, 0.814)	
        (e, 10.814) +- (0, 3.025)	
        (f, 21.864) +- (0, 13.603)	
        (g, 363.142) +- (0, 16.089)	
        (h, 17.900) +- (0, 3.399)	
        (i, 13.320) +- (0, 3.832)	
        (j, 279.362) +- (0, 87.451)	
        (k, 39.628) +- (0, 11.862)	
        (l, 72.480) +- (0, 27.110)	
        (m, 110.511) +- (0, 36.613)	
        (n, 124.520) +- (0, 87.167)	
		};
    \addplot+[
        error bars/.cd,
        y dir=both,
        y explicit
    ] coordinates {
        (a, 0.043) +- (0, 0.001)	
        (b, 1.225) +- (0, 0.020)	
        (c, 3.169) +- (0, 0.202)	
        (d, 3.041) +- (0, 0.351)	
        (e, 12.944) +- (0, 1.953)	
        (f, 20.671) +- (0, 3.526)	
        (g, 738.673) +- (0, 59.774)	
        (h, 8.961) +- (0, 1.078)	
        (i, 95.186) +- (0, 15.603)	
        (j, 594.911) +- (0, 278.741)	
        (k, 29.790) +- (0, 7.743)	
        (l, 176.663) +- (0, 66.081)	
        (m, 497.072) +- (0, 143.320)	
        (n, 634.630) +- (0, 111.511)	
		};
\end{vsplot}
\caption{
		Average runtimes with standard deviations of the constraining, relaxing and binary search strategies to solve the pebbling problem with naive \pdr, \spacer and \ipdr. 
}
 \label{fig:all}
\end{figure}

\paragraph{Results for Pebbling}
	\autoref{fig:pebbling:constrain:all} and \autoref{fig:all} show the benchmark results for the
	constraining (\cipdr), relaxing (\ripdr) and binary search  versions of \ipdr on the pebbling problem.
	In over half of the benchmarks, constraining \ipdr achieves a
	speedup of around roughly $50\%$ (a factor two) compared to naive \pdr and  \spacer. 
	For larger benchmarks, however, \spacer has a clear advantage, reducing runtimes by a factor five, a trend that persists for the relaxing and binary search strategies as well.
	This speedup range  falls
	outside of the standard deviation of naive \ipdr and \spacer.
Relaxing \ipdr appears to achieve little advantage over the other methods, a result contrary to what we observed for the Peterson protocol.

\renewcommand\hangnote{}
\renewcommand{\arraystretch}{1.5}

\begin{figure}[b]
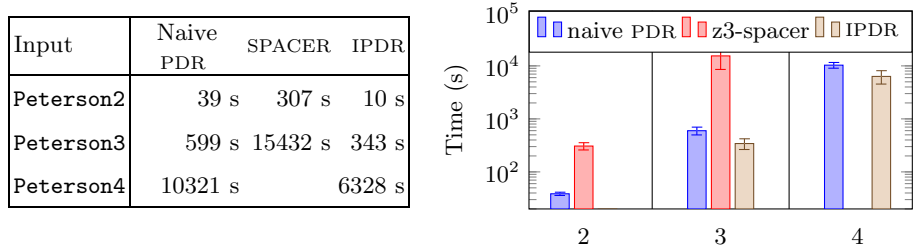

\small
\begin{minipage}{5.5cm}
\vspace{-11em}
\begin{tabular}{|l|rrr|}
  \hline
      Input 
    & \pbox{1.4cm}{\vspace{.2\baselineskip}\centering
      Naive \pdr \vspace{.2\baselineskip}} 
    & \spacer 
    & \ipdr 
\\
     
  \hline 
     
  \ttfunction{Peterson2}%
    \hangnote 						 & 39 s     &   307 s & 10 s  \\
  \ttfunction{Peterson3}%
    \hangnote 						 & 599 s    & 15432 s & 343 s \\
  \ttfunction{Peterson4} & 10321 s  & 				    & 6328 s  \\
  \hline
\end{tabular}
\end{minipage}
\begin{vsplotpeter}{naive \pdr, z3-spacer, \ipdr}{
    2, 3, 4
}
    \addplot+[
        error bars/.cd,
        y dir=both,
        y explicit, 
    ] coordinates {
    	(2, 38.654) +- (0, 2.729)	
      (3, 598.523) +- (0, 100.434)	
      (4, 10321.160) +- (0, 1279.712)	
    };
    
    \addplot+[
        error bars/.cd,
        y dir=both,
        y explicit,
    ] coordinates {
    	(2, 307.348) +- (0, 48.051)	
      (3, 15431.695) +- (0, 6862.066)	
    };
    
    \addplot+[
        error bars/.cd,
        y dir=both,
        y explicit
    ] coordinates {
			(2, 10.144) +- (0, 1.371)	
      (3, 342.747) +- (0, 78.325)	
      (4, 6328.118) +- (0, 1788.829)	
		};
\end{vsplotpeter} 
\vspace{-.8em}
\caption{
		Average runtimes with standard deviation for Peterson's protocol for
    naive~\pdr, \spacer and relaxing \ipdr.
  For \ttfunction{Peterson4}, \spacer timed out.
}
 \label{fig:peter:relax:all}
\end{figure}

Finally, we see that speedups of the constraining version are preserved by binary search, but not improved. Internal statistics clearly show that a portion of the incremental runs complete extremely fast, because they are either over- or under-constrained. However, when the binary search approaches the optimal pebble count, the runs become expensive (a well-known threshold behavior in SAT solving~\cite{heule2016solving,heule2017science,heule2018schur}). The binary search is unable to reduce the number of expensive incremental runs close to the optimal number of pebbles, because it is now approached from both above and below.

\paragraph{Statistics}
\autoref{app:stats}
 presents internal statistics of the \ipdr algorithm on the  pebbling of the \texttt{ham7tc} and \texttt{5bitadder} circuits and the context-bounded  \texttt{Peterson4} protocol.
We selected the two circuits because constraining \ipdr yields a speedup on both, while relaxing \ipdr only achieves speedup on the  latter. 
 The statistics gathered are the number of counterexamples to induction (CTIs)~\cite{bradley2011sat}, the number of proof obligations queued,  the number of SAT solver calls and the number of cubes/clauses that can be carried over to the next (relaxing) increment (see \autoref{alg:ipdr:relax-prep}). 
Aside from the counts, we also recorded the time taken to perform these actions. These results are plotted per incremental \pdr run.
	We recorded statistics for both constraining and relaxing runs.

These measurements reveal that constraining \ipdr is able to reduce by a factor six the number of counterexamples to induction for incremental instances compared to naive \pdr. 
This factor tends to reduce when approaching the optimal number of pebbles (threshold behavior that was observed elsewhere~\cite{heule2016solving,heule2017science,heule2018schur}).
Nonetheless, the result is that \ipdr is consistently about as fast for increment $i$  as naive \pdr for increment $i +1$.
For relaxing \ipdr, we do not always observe this behavior (also not for instances with positive speedups), but \ripdr is able to copy 60\% of the blocked clauses between increments for both tested circuits. 
However, the copying process is expensive for the \texttt{ham7tc} circuit, which negates any performance benefits for the subsequent incremental \ipdr run.

\renewcommand{\arraystretch}{.994}
	\begin{figure}[t!]
\small
\centering
\begin{tabular}{|c|cl|rrr|rr|}
	\hline
			&
			& Input 
			& \pbox{1.4cm}{\vspace{.2\baselineskip}\centering
					Naive \pdr \vspace{.2\baselineskip}} 
			& \spacer 
			& \ipdr 
			& \pbox{1.3cm}{\vspace{.2\baselineskip}\centering
					Speedup vs. naive\vspace{.2\baselineskip}}
			& \pbox{1.7cm}{\vspace{.2\baselineskip}\centering
					Speedup vs. \spacer\vspace{.2\baselineskip}}\\

		\hline
\multirow{15}{1em}{\turner{90}{\textbf{{Constraining strategy}}}} 
		&a & \texttt{ham3tc}                & 0.031 s    & 0.281 s   & 0.031 s    & 1\%    & 89\%   \\
		&b & \texttt{mod5d1}                & 1.942 s    & 2.850 s   & 1.067 s    & 45\%   & 63\%   \\
		&c & \texttt{gf2\^{}3mult\_11\_47}  & 2.545 s    & 2.670 s   & 2.751 s    & \slower -8\%   & \slower -3\%   \\
		&d & \texttt{nth\_prime4\_inc\_d1}  & 3.428 s    & 4.098 s   & 2.878 s    & 16\%  &  30\%   \\
		&e & \texttt{4b15g\_1}\hangnote     & 34.711 s   & 13.225 s  & 11.513 s   & 67\%   & 13\%   \\
		&f & \texttt{4\_49tc1}              & 55.970 s   & 37.466 s  & 18.761 s   & 66\%   & 50\%   \\
		&g & \texttt{5mod5tc}               & 989.706 s  & 377.234 s & 748.969 s  & 24\%   & \slower -99\%  \\
		&h & \texttt{hwb4tc}                & 30.292 s   & 30.153 s  & 9.865 s    & 67\%   & 67\%   \\
		&i & \texttt{gf2\^{}4mult\_19\_83}  & 19.618 s   & 25.678 s  & 111.350 s  & \slower -468\% & \slower -334\% \\
		&j & \texttt{rd73d2} \hangnote      & 836.655 s  & 351.181 s & 499.460 s  & 40\%   & \slower -42\%  \\
		&k & \texttt{mod5adders}\hangnote   & 57.610 s   & 57.699 s  & 34.781 s   & 40\%   & 40\%   \\
		&l & \texttt{ham7tc}                & 386.719 s  & 90.823 s  & 170.244 s  & 56\%   & \slower -87\%  \\
		&m & \texttt{5bitadder}             & 1396.920 s & 341.008 s & 964.889 s  & 31\%   & \slower -183\% \\
		&n & \texttt{gf2\^{}5mult\_29\_129} & 760.006 s  & \phantom{1}242.379 s & 1243.961 s & \slower -64\%  & \slower -413\% \\
		\hline
\end{tabular}
\end{figure}

	\begin{figure}[p!]
\vspace{-.5em}
\small
\centering
\begin{tabular}{|c|cl|rrr|rr|}

\cull{
	\hline
			&& Input 
			& \pbox{1.4cm}{\vspace{.2\baselineskip}\centering
					Naive \pdr \vspace{.2\baselineskip}} 
			& \spacer 
			& \ipdr 
			& \pbox{1.3cm}{\vspace{.2\baselineskip}\centering
					Speedup vs. naive\vspace{.2\baselineskip}}
			& \pbox{1.7cm}{\vspace{.2\baselineskip}\centering
					Speedup vs. \spacer\vspace{.2\baselineskip}}\\
}{
			&& \hphantom{Input}\vspace{-2.3em}
			& \hphantom{\pbox{1.4cm}{\vspace{.2\baselineskip}\centering
					Naive \pdr \vspace{.2\baselineskip}}}
			& \hphantom{\spacer }
			& \hphantom{\ipdr }
			& \hphantom{\pbox{1.3cm}{\vspace{.2\baselineskip}\centering
					Speedup vs. naive\vspace{.2\baselineskip}}}
			& \hphantom{\pbox{1.7cm}{\vspace{.2\baselineskip}\centering
					Speedup vs. \spacer\vspace{.2\baselineskip}}}\\
}
		\hline

\multirow{14}{1em}{\turner{90}{\textbf{Relaxing strategy}}} 
		&a & \texttt{ham3tc}                & 0.041 s   & 0.144 s   & 0.039 s   & 6\%   & 73\%   \\
		&b & \texttt{mod5d1}                & 0.832 s   & 1.413 s   & 0.646 s   & 22\%  & 54\%   \\
		&c & \texttt{gf2\^{}3mult\_11\_47}  & 1.323 s   & 1.416 s   & 1.159 s   & 12\%  & 18\%   \\
		&d & \texttt{nth\_prime4\_inc\_d1}\hangnote  & 2.647 s   & 2.433 s   & 2.806 s   & \slower \slower -6\%  & \slower -15\%  \\
		&e & \texttt{4b15g\_1}              & 16.901 s  & 8.773 s   & 19.870 s  & \slower -18\% & \slower -126\% \\
		&f & \texttt{4\_49tc1}              & 43.018 s  & 19.833 s  & 63.133 s  & \slower -47\% & \slower -218\% \\
		&g & \texttt{5mod5tc}               & 845.274 s & 324.640 s & 747.978 s & 12\%  & \slower -130\% \\
		&h & \texttt{hwb4tc}                & 15.325 s  & 14.018 s  & 22.853 s  & \slower -49\% & \slower -63\%  \\
		&i & \texttt{gf2\^{}4mult\_19\_83}  & 5.755 s   & 8.441 s   & 5.366 s   & 7\%   & 36\%   \\
		&j & \texttt{rd73d2}\hangnote & 634.672 s & \phantom{1}202.320 s & 654.798 s & \slower -3\%  & \slower -224\% \\
		&k & \texttt{mod5adders}            & 45.037 s  & 38.639 s  & 28.258 s  & 37\%  & 27\%   \\
		&l & \texttt{ham7tc}                & 151.867 s & 53.712 s  & 211.390 s & \slower -39\% & \slower -294\% \\
		&m & \texttt{5bitadder}             & 639.886 s & 91.152 s  & 474.993 s & 26\%  & \slower -421\% \\
		&n & \texttt{gf2\^{}5mult\_29\_129} & \phantom{1}211.999 s & 64.588 s  & \phantom{1}255.076 s & \slower -20\% & \slower -295\% \\ \hline
\end{tabular}
\end{figure}

	\begin{figure}[p!]
\cull{}{\vspace{-.5em}}

\small
\centering
\begin{tabular}{|c|ll|rrr|rr|}

\cull{ 
  \hline
  &
  & Input 
  & \pbox{1.4cm}{\vspace{.2\baselineskip}\centering
    Naive \pdr \vspace{.2\baselineskip}} 
  & \spacer 
  & \ipdr 
  & \pbox{1.3cm}{\vspace{.2\baselineskip}\centering
    Speedup vs. naive\vspace{.2\baselineskip}}
  & \pbox{1.7cm}{\vspace{.2\baselineskip}\centering
    Speedup vs. \spacer\vspace{.2\baselineskip}}\\
}{
			&& \hphantom{Input}\vspace{-2.3em}
			& \hphantom{\pbox{1.4cm}{\vspace{.2\baselineskip}\centering
					Naive \pdr \vspace{.2\baselineskip}}}
			& \hphantom{\spacer }
			& \hphantom{\pdr }
			& \hphantom{\pbox{1.3cm}{\vspace{.2\baselineskip}\centering
					Speedup vs. naive\vspace{.2\baselineskip}}}
			& \hphantom{\pbox{1.7cm}{\vspace{.2\baselineskip}\centering
					Speedup vs. \spacer\vspace{.2\baselineskip}}}\\
}

  \hline
 \multirow{14}{1em}{\turner{90}{\textbf{Binary search strategy}}}
  &a & \texttt{ham3tc}                & 0.135 s   & 0.284 s   & 0.043 s   & 68\%           & 85\%           \\
  &b & \texttt{mod5d1}                & 1.297 s   & 1.573 s   & 1.225 s   & 6\%            & 22\%           \\
  &c & \texttt{gf2\^{}3mult\_11\_47}  & 2.124 s   & 1.861 s   & 3.169 s   & \slower -49\%  & \slower -70\%  \\
  &d & \texttt{nth\_prime4\_inc\_d1}  & 3.776 s   & 4.239 s   & 3.041 s   & 19\%           & 28\%           \\
  &e & \texttt{4b15g\_1}\hangnote     & 28.433 s  & 10.814 s  & 12.944 s  & 54\%           & \slower -20\%  \\
  &f & \texttt{4\_49tc1}              & 47.795 s  & 21.864 s  & 20.671 s  & 57\%           & 5\%            \\
  &g & \texttt{5mod5tc}               & 938.345 s & 363.142 s & 738.673 s & 21\%           & \slower -103\% \\
  &h & \texttt{hwb4tc}                & 19.572 s  & 17.900 s  & 8.961 s   & 54\%           & 50\%           \\
  &i & \texttt{gf2\^{}4mult\_19\_83}  & 13.916 s  & 13.320 s  & 95.186 s  & \slower -584\% & \slower -615\% \\
  &j & \texttt{rd73d2}                & 714.732 s & 279.362 s & 594.911 s & 17\%           & \slower -113\% \\
  &k & \texttt{mod5adders}\hangnote   & 45.037 s  & 39.628 s  & 29.790 s  & 34\%           & 25\%           \\
  &l & \texttt{ham7tc}                & 289.942 s & 72.480 s  & 176.663 s & 39\%           & \slower -144\% \\
  &m & \texttt{5bitadder}             & 941.210 s & 110.511 s & 497.072 s & 47\%           & \slower -350\% \\
  &n & \texttt{gf2\^{}5mult\_29\_129} & \phantom{1}423.146 s & \phantom{1}124.520 s & \phantom{1}634.630 s & \slower -50\%  & \slower -410\% \\ \hline
\end{tabular}
\caption{
		Average runtimes the constraining, relaxing and binary search strategies to solve the pebbling problem with naive \pdr, \spacer and \ipdr. Standard deviations are plotted in \autoref{fig:all}.
		Speedup denotes the percentage runtime decrease or increase (gray) achieved by \ipdr:
		$1- \frac{\text{time(\ipdr)}}{\text{time(other)}}$.
}
\label{fig:pebbling:constrain:all}
\label{fig:pebbling:relax:all}
\label{fig:pebbling:binary:all}
\end{figure}

\section{Conclusions}
\label{sec:conclusion}

We introduced Incremental Property Directed Reachability (\ipdr), which harnesses the strength of incremental SAT solvers to prove correctness of parameterized systems.
Since \pdr does not use unrolling like in bounded model checking, we identified other structural parameters for \ipdr to exploit: a bound on the number of interleavings in the parallel program and the maximum number of pebbles used to solve the pebbling game optimization problem.

With an open source implementation of \ipdr, we demonstrated that the incremental approach can optimize pebbling games and model check  parallel programs faster than \spacer~\cite{z3,spacer}.
We observed that constraining \ipdr outperforms its relaxing counterpart, while surprisingly binary search adds little improvements over the constraining approach. We can attribute the latter to the hardness of the incremental \pdr runs close to optimum: While under- and over-constrained systems are ofter solved in negligible time, those closer to the optimum take the most time (a well-known behavior~\cite{heule2016solving,heule2017science,heule2018schur}). Consequently, the runtime of binary search is dominated by the instances around the optimum where the search often concentrates.

Finally, internal counters from the \ipdr implementations reveal that ample of information is reused in both relaxing and constraining incremental runs. 
We therefore expect that further research on different parameters and other problem instances will reveal more benefits of the \ipdr approach.

\bibliographystyle{plain}
\bibliography{lit}

\appendix
\newpage
\section{Statistics from \ipdr Runs}
\label{app:stats}

This appendix presents statistics about the internals of the \ipdr algorithm. The statistics gathered are the number of counterexamples to induction (CTIs)~\cite{bradley2007checking}, the number of obligations queued,  the number of SAT solver calls and the number of cubes/clauses that can be carried over to the next (relaxing) increment (see \autoref{alg:ipdr:relax-prep}). Aside from the counts, we also recorded the time taken to perform these actions. These results are plotted per incremental \pdr run. (Note that some incremental runs are not recorded, since \pdr can return a shorter counterexample so that we may skip multiple instances at once.)
	We recorded statistics for both constraining and relaxing runs.

\begin{figure}[ht]
    \centering
    \begin{tabular}
        {|r|rr|r|}
        \hline
        & \cipdr & Naive \cipdr & Improvement \\\hline
		avg time & 170.244 s & 386.719 s & 55.98 \% \\
		std dev time & 33.414 s & 62.920 s &  \\
		avg incremental time & 0.594 s &  &  \\
		std dev incremental time & 0.044 s &  &  \\
		max inv constraint & 9 & 9 &  \\
		min inv level & 18 & 9 & 0.00 \% \\
		min strat marked & 10 & 10 &  \\
        min strat length & 25 & 26 & 3.85 \% \\\hline
    \end{tabular}

	\vspace{1em}
	
\begin{subfigure}{\textwidth}
	\begin{tikzpicture}
	    \begin{axis}
     [
	        xtick=data,
	        xtick style={draw=none},
	        x axis line style={draw=none},
	        minor tick num=1,
	        height=6cm,
	        width=\textwidth,
	        enlarge x limits=0.05,
	        enlarge y limits={upper=0},
	        xlabel={Maximum Number of Pebbles},
	        xticklabels={9,10,11,12,13,14,15,16,17,23},
	        ybar,
	        axis y line*=left,
	        bar width=7pt,
	        legend style={at={(0.5,1.13)}, anchor=north,legend columns=-1},
	        ylabel={No. ctis},
	        x dir = reverse,
	        fill between/on layer={main},
     ]
	        \legend{\cipdr cti-count, naive cti-count}
					\pgfplotstableread{
	            x y err
	            9 0.3 0.09486832980505137
	            10 1.4 0.12649110640673514
	            11 1 0
	            12 1 0
	            13 1 0
	            14 1 0
	            15 1 0
	            16 1 0
	            17 1 0
	            19 6 0
	        }\cticount
	        \errorbarplot{blue}{\cticount};
					\pgfplotstableread{
	            x y err
	            9 4 0
	            10 6.2 1.0411532067856297
	            11 7.1 1.0397114984456024
	            12 6.5 0.9082951062292475
	            13 6 0
	            14 6 0
	            15 6 0
	            16 6 0
	            17 6 0
	            19 6 0
	        }\ncticount
	        \errorbarplot{red}{\ncticount};
	        
        	\defxcrunch{0.11}{0.16}
         \end{axis}
     
 		\drawxcrunch

	    \begin{axis}
	        [
	        ymode=log,
	        log origin=infty,
	        xtick=data,
	        xtick style={draw=none},
	        x axis line style={draw=none},
	        minor tick num=1,
	        height=6cm,
	        width=\textwidth,
	        enlarge x limits=0.05,
	        enlarge y limits={upper=0},
	        axis y line*=right,
	        legend style={at={(0.5,1.0)}, anchor=north,legend columns=-1},
	        xticklabels={,,},
	        ylabel={Time (s)},
	        x dir = reverse,
	        fill between/on layer={main},
	        ]
	        \legend{\cipdr cti-time, naive cti-time}
			    \pgfplotstableread{
	            x y err
	            9 1.2884875408 1.4163873999863286
	            10 106.358287217 27.658799733106907
	            11 36.190334611599994 24.527116723681885
	            12 12.043066895900001 3.6211082402196175
	            13 6.210486761 1.724949667620505
	            14 2.0798271135 1.0106151242017232
	            15 1.2472325008571428 0.7406256415384112
	            16 1.2118622063333333 0.4726961066133389
	            17 1.563704145 0
	            19 4.0869381753 0.19456246266195956
	        }\ctitime
	        \dotplot{blue}{\ctitime};
	        \addplot [name path=upper,draw=none]
	        table[x=x,y expr=\thisrow{y}+\thisrow{err}] {\ctitime};
	        \addplot [name path=lower,draw=none] 
	        table[x=x,y expr=\thisrow{y}-\thisrow{err}] {\ctitime};
	        \filldevplot{blue}
			\pgfplotstableread{
	            x y err
	            9 3.6242223535999996 0.018413229722648895
	            10 189.87885999079998 28.629797816176907
	            11 112.64717464190001 40.58501038470905
	            12 45.1601882996 13.753186424023555
	            13 14.863202483699999 3.8061174979311985
	            14 8.564519906799998 1.4145684397135485
	            15 6.747898380999999 0.5819410110890614
	            16 6.9705064640000005 0.1823504259744781
	            17 6.153664137 0
	            19 4.0933943409 0.19540634219906944
	        }\ncticount
	        \dotplot{red}{\ncticount};
	        \addplot [name path=upper,draw=none]
	        	table[x=x,y expr=\thisrow{y}+\thisrow{err}] {\ncticount};
	        \addplot [name path=lower,draw=none] 
	        	table[x=x,y expr=\thisrow{y}-\thisrow{err}] {\ncticount};
	        \linedevplot{red}
	        
	    \end{axis}
	    
	\end{tikzpicture}
	\caption{
		A comparison between the number of counterexamples found in the major
		iteration (cti-count) on the left axis and the total runtime (cti-time)
		on the right axis.
	}
	\label{fig:pebbling:constrain:ham7:cti}
	\end{subfigure}
	
	\vspace{1em}
	
	\begin{subfigure}{\textwidth}
		\centering
	\begin{tikzpicture}		
		\begin{axis}
		[
			xtick=data,
			xtick style={draw=none},
			minor tick num=1,
			height=4cm,
			width=0.8\textwidth,
			enlarge x limits=0.05,
			enlarge y limits={upper=0},
			xlabel={Maximum Number of Pebbles},
			legend style={at={(0.5,1.23)}, anchor=north,legend columns=-1},
			ylabel={Time (s)},
			x dir = reverse,
			fill between/on layer={main},
		]
			\legend{\cipdr inc-time, naive inc-time}
			\pgfplotstableread{
				x y err
			    9 0.26132231679999995 0.03399652673484897
			    10 0.1215890968 0.0188224094715695
			    11 0.07217729329999999 0.007775473612871219
			    12 0.04853685640000001 0.00445019304784721
			    13 0.0336318478 0.0027108322124754537
			    14 0.028536875499999996 0.0034852225626870728
			    15 0.02628886714285714 0.0018837199875151615
			    16 0.024544915333333334 0.0024857351622102116
			    17 0.025080067 0
			}\inctime
			\dotplot{blue}{\inctime};
			\addplot [name path=upper,draw=none]
				table[x=x,y expr=\thisrow{y}+\thisrow{err}] {\inctime};
			\addplot [name path=lower,draw=none] 
				table[x=x,y expr=\thisrow{y}-\thisrow{err}] {\inctime};
			\filldevplot{blue}
			\pgfplotstableread{
				x y err
				9 0.0662037737 0.0009982932135989967
				10 0.06513288390000001 0.0007058335541635077
				11 0.064573517 0.0009061616199195369
				12 0.063899825 0.0007116947427317436
				13 0.0637975121 0.001095456718934567
				14 0.064155989 0.0017916209965402826
				15 0.06311204671428572 0.00015336877614961894
				16 0.06293949433333333 0.00015765172853546926
				17 0.062762295 0
			}\ninctime
		\dotplot{red}{\ninctime};
		\addplot [name path=upper,draw=none]
			table[x=x,y expr=\thisrow{y}+\thisrow{err}] {\ninctime};
		\addplot [name path=lower,draw=none] 
			table[x=x,y expr=\thisrow{y}-\thisrow{err}] {\ninctime};
		\linedevplot{red}
		\end{axis}
	\end{tikzpicture}
	\caption{
		The amount of the runtime that was spent on copying clauses to the
		constrained instance denoted on the x-axis. This number is already included
		in the times of the upper graph. 
	}
	\label{fig:pebbling:constrain:ham7:inc}
	\end{subfigure}

    \caption{
	    Statistics for the constraining \texttt{ham7tc} experiment,
	    averaged over 10 repetitions.
	    On constraint 23, the algorithm found a trace with 18 pebbles, so it could continue with constraint 17 afterwards.
    }
    \label{fig:pebbling:constrain:ham7}
\end{figure}
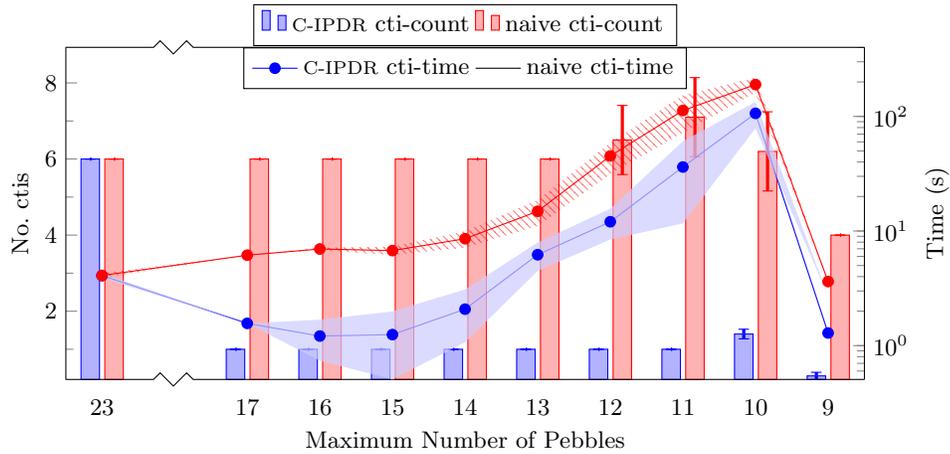
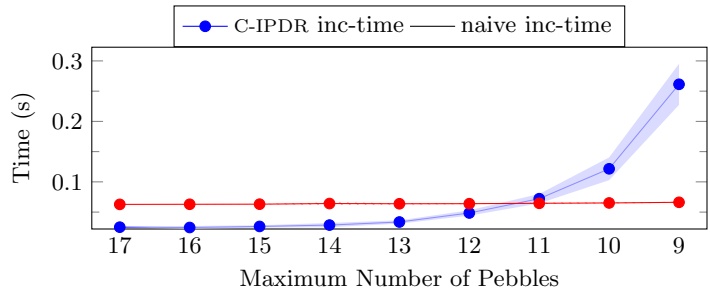

	\paragraph{Constraining ham7tc}
	\autoref{fig:pebbling:constrain:ham7} and
	\autoref{fig:pebbling:constrain:ham7:oblsat}	
		show detailed statistics from the  \texttt{ham7tc} experiment with \cipdr, which performed well in the constraining setting.
		\autoref{fig:pebbling:constrain:ham7:cti} shows that \cipdr decreases the number of CTIs   considered in all incremental \ipdr iterations after the first, with a corresponding decrease in runtime. The graph shows that the \cipdr and naive versions have a similar exponential scaling in runtimes as the iterations progress, but that \cipdr iterations are consistently as fast as \pdr on the previous incremental instance. 
		\autoref{fig:pebbling:constrain:ham7:oblsat} shows a similar relation for the number of handled obligations and \sat-queries.
		Finally, the runtime spent on copying blocking clauses between iterations is negligible as \autoref{fig:pebbling:constrain:ham7:inc} shows.

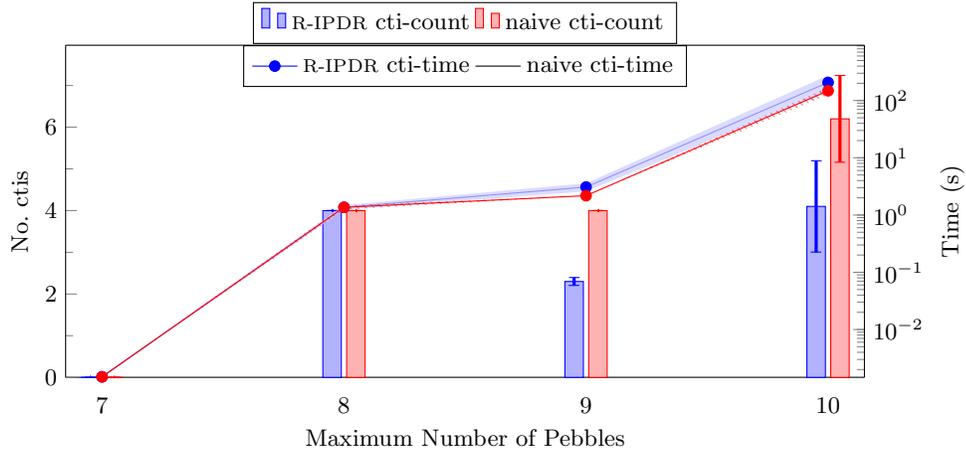
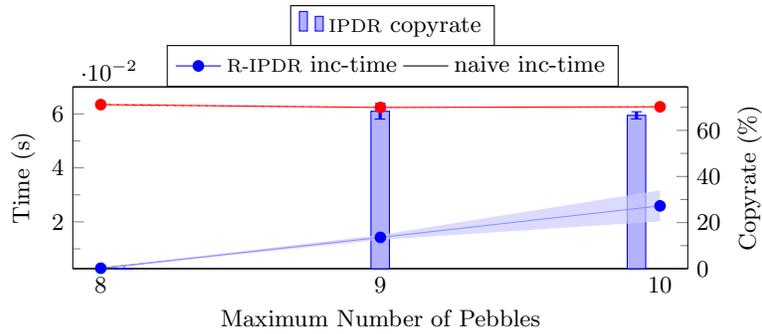
\begin{figure}[ht]
    \centering
    \begin{tabular}
        {|r|rr|r|}
		\hline
 	& \ripdr & Naive \ripdr & Improvement \\\hline
		avg time & 211.390 s & 151.867 s & -39.19 \% \\
		std dev time & 67.712 s & 22.490 s &  \\
		avg incremental time & 0.043 s &  &  \\
		std dev incremental time & 0.006 s &  &  \\
		max inv constraint & 9 & 9 &  \\
		min strat marked & 10 & 10 &  \\
		min strat length & 27 & 26 & -3.85 \% \\\hline
    \end{tabular}

		\vspace{1em}

\begin{subfigure}{\textwidth}
	\begin{tikzpicture}
	    \begin{axis}
     [
	        xtick=data,
	        xtick style={draw=none},
	        x axis line style={draw=none},
	        minor tick num=1,
	        height=6cm,
	        width=\textwidth,
	        enlarge x limits=0.05,
	        enlarge y limits={upper=0},
	        xlabel={Maximum Number of Pebbles},
	        ybar,
	        axis y line*=left,
	        bar width=7pt,
	        legend style={at={(0.5,1.13)}, anchor=north,legend columns=-1},
	        ylabel={No. ctis},
	        fill between/on layer={main},
     ]
	        \legend{\ripdr cti-count, naive cti-count}
					\pgfplotstableread{
							x y err
							7 0 0
							8 4 0
							9 2.3 0.09486832980505133
							10 4.1 1.0959014554237987
	        }\cticount
	        \errorbarplot{blue}{\cticount};
					\pgfplotstableread{
							x y err
							7 0 0
							8 4 0
							9 4 0
							10 6.2 1.0411532067856297
	        }\ncticount
	        \errorbarplot{red}{\ncticount};
			\end{axis}

	    \begin{axis}
			[
	        ymode=log,
	        log origin=infty,
	        xtick=data,
	        xtick style={draw=none},
	        minor tick num=1,
	        height=6cm,
	        width=\textwidth,
	        enlarge x limits=0.05,
	        enlarge y limits={upper=0},
	        axis y line*=right,
	        legend style={at={(0.5,1.0)}, anchor=north,legend columns=-1},
	        xticklabels={,,},
	        ylabel={Time (s)},
	        fill between/on layer={main},
			]
	        \legend{\ripdr cti-time, naive cti-time}
					\pgfplotstableread{
							x y err
							7 0.0014916292999999998 2.834963302425625e-05
							8 1.3625377523000002 0.12171274773235408
							9 3.0815468609000005 0.4999340957941288
							10 206.90099932439998 67.6943118030946
	        }\ctitime
	        \dotplot{blue}{\ctitime};
	        \addplot [name path=upper,draw=none]
						table[x=x,y expr=\thisrow{y}+\thisrow{err}] {\ctitime};
	        \addplot [name path=lower,draw=none] 
						table[x=x,y expr=\thisrow{y}-\thisrow{err}] {\ctitime};
	        \filldevplot{blue}
					\pgfplotstableread{
             	x y err
	            7 0.0014905496 2.2417217834512847e-05
	            8 1.3624095955 0.12265441338770981
	            9 2.1766463017999995 0.0134865784371539
	            10 148.1380169876 22.492937123032096
	        }\nctitime
	        \dotplot{red}{\nctitime};
	        \addplot [name path=upper,draw=none]
	        	table[x=x,y expr=\thisrow{y}+\thisrow{err}] {\nctitime};
	        \addplot [name path=lower,draw=none] 
	        	table[x=x,y expr=\thisrow{y}-\thisrow{err}] {\nctitime};
	        \linedevplot{red}
	        
	    \end{axis}
	    
	\end{tikzpicture}
	\caption{
		A comparison between the number of counterexamples found in the major
		iteration (cti-count) on the left axis and the total runtime (cti-time)
		on the right axis.
	}
	\label{fig:pebbling:relax:ham7:cti}
	\end{subfigure}

		\vspace{1em}
	
	\begin{subfigure}{\textwidth}
		\centering
	\begin{tikzpicture}		
		\begin{axis}
			[
				ymin=0,
				xtick=data,
				xtick style={draw=none},
				minor tick num=1,
				height=4cm,
				width=0.8\textwidth,
				enlarge x limits=0.1,
				enlarge y limits={upper=0},
				ybar,
				axis y line*=right,
				xticklabels={,,},
				bar width=7pt,
				legend style={at={(0.52,1.45)}, anchor=north,legend columns=-1},
				ylabel={Copyrate (\%)},
				fill between/on layer={main},
			]
			\legend{\ipdr copyrate}
			\pgfplotstableread{
					x y err
					8 0 0
					9 68.27586206896551 3.3786065417698965
					10 66.51378033730974 1.5513039743986192
			}\incperc
			\errorbarplot{blue}{\incperc};
		\end{axis}
	
		\begin{axis}
		[
			xtick=data,
			xtick style={draw=none},
			minor tick num=1,
			height=4cm,
			width=0.8\textwidth,
			enlarge x limits=0.05,
			enlarge y limits={upper=0},
			xlabel={Maximum Number of Pebbles},
			axis y line*=left,
			legend style={at={(0.52,1.23)}, anchor=north,legend columns=-1},
			ylabel={Time (s)},
			fill between/on layer={main},
		]
			\legend{\ripdr inc-time, naive inc-time}
			\pgfplotstableread{
				x y err
				8 0.0028068212 0.00014898600661995078
				9 0.014244686599999997 0.0009128519472177511
				10 0.025914295999999996 0.005760409145567388
			}\inctime
			\dotplot{blue}{\inctime};
			\addplot [name path=upper,draw=none]
				table[x=x,y expr=\thisrow{y}+\thisrow{err}] {\inctime};
			\addplot [name path=lower,draw=none] 
				table[x=x,y expr=\thisrow{y}-\thisrow{err}] {\inctime};
			\filldevplot{blue}
			\pgfplotstableread{
			    x y err
					8 0.0634388488 0.0004648420572492997
					9 0.06239420130000001 0.0001754093652055396
					10 0.062633937 0.000277725819716858
			}\ninctime
		\dotplot{red}{\ninctime};
		\addplot [name path=upper,draw=none]
			table[x=x,y expr=\thisrow{y}+\thisrow{err}] {\ninctime};
		\addplot [name path=lower,draw=none] 
			table[x=x,y expr=\thisrow{y}-\thisrow{err}] {\ninctime};
		\linedevplot{red}
		\end{axis}
	\end{tikzpicture}
	\caption{
		The amount of the runtime that was spent on copying clauses to the
		relaxed instance denoted on the x-axis. This number is already included
		in the times of the upper graph. 
	}
	\label{fig:pebbling:relax:ham7:copy}
	\end{subfigure}

    \caption{
	    Statistics for the relaxing \texttt{ham7tc} experiment,
	    averaged over 10 repetitions.
    }\label{fig:pebbling:relax:ham7}
\end{figure}

	\paragraph{Relaxing ham7tc}
	\autoref{fig:pebbling:relax:ham7} and 
	\autoref{fig:pebbling:relax:ham7:oblsat} provide the same information for relaxing \ipdr on the \texttt{ham7tc} circuit.
	\autoref{fig:pebbling:relax:ham7:cti} shows that \cipdr  decreases the number of CTIs considered in  incremental \ipdr iterations 8 and 9. A decrease in runtime is however absent. 
		\autoref{fig:pebbling:relax:ham7:oblsat} shows a similarly meagre benefits of the relaxing approach in the number of handled obligations and \sat-queries.
	\autoref{fig:pebbling:relax:ham7:copy} reveals that  \autoref{alg:ipdr:relax-prep} is able to copy over 60\% of the blocking clauses to the next increment but requires relatively much time to do so
		compared to e.g. CTI handling~\autoref{fig:pebbling:relax:ham7:cti}.

\begin{figure}[ht]
\centering
\begin{tabular}
{|r|rr|r|}
\hline
 	& \ripdr & Naive \ripdr & Improvement \\\hline
	avg time & 474.993 s & 639.886 s & 25.77 \% \\
	std dev time & 171.428 s & 150.129 s &  \\
	avg incremental time & 0.149 s &  &  \\
	std dev incremental time & 0.006 s &  &  \\
	max inv constraint & 8 & 8 &  \\
	min strat marked & 9 & 9 &  \\
	min strat length & 38 & 36 & -5.56 \% \\\hline
\end{tabular}

		\vspace{1em}

\begin{subfigure}{\textwidth}
	\begin{tikzpicture}
	    \begin{axis}
			 [
			 			ymin=0,
						xtick=data,
						xtick style={draw=none},
						x axis line style={draw=none},
						minor tick num=1,
						height=6cm,
						width=\textwidth,
						enlarge x limits=0.05,
						enlarge y limits={upper=0},
						xlabel={Maximum Number of Pebbles},
						ybar,
						axis y line*=left,
						bar width=7pt,
						legend style={at={(0.5,1.13)}, anchor=north,legend columns=-1},
						ylabel={No. ctis},
						fill between/on layer={main},
			 ]
	        \legend{\ripdr cti-count, naive cti-count}
					\pgfplotstableread{
							x y err
							6 6 0
							7 2 0
							8 3.3 0.2213594362117866
							9 2.7 0.09486832980505133
					}\cticount
	        \errorbarplot{blue}{\cticount};
					\pgfplotstableread{
							x y err
							6 6 0
							7 6 0
							8 7 0
							9 7.1 0.03162277660168368
					}\ncticount
	        \errorbarplot{red}{\ncticount};
			\end{axis}

	    \begin{axis}
			[
	        ymode=log,
	        log origin=infty,
	        xtick=data,
	        xtick style={draw=none},
	        minor tick num=1,
	        height=6cm,
	        width=\textwidth,
	        enlarge x limits=0.05,
	        enlarge y limits={upper=0},
	        axis y line*=right,
	        legend style={at={(0.5,1.0)}, anchor=north,legend columns=-1},
	        xticklabels={,,},
	        ylabel={Time (s)},
	        fill between/on layer={main},
			]
	        \legend{\ripdr cti-time, naive cti-time}
					\pgfplotstableread{
							x y err
							6 2.2219050884000007 0.032690346066451666
							7 1.6642746847 0.019172573457669505
							8 50.69399441650001 2.8613003166469673
							9 420.26344129930004 171.8547871248775
					}\ctitime
	        \dotplot{blue}{\ctitime};
	        \addplot [name path=upper,draw=none]
						table[x=x,y expr=\thisrow{y}+\thisrow{err}] {\ctitime};
	        \addplot [name path=lower,draw=none] 
						table[x=x,y expr=\thisrow{y}-\thisrow{err}] {\ctitime};
	        \filldevplot{blue}
					\pgfplotstableread{
             	x y err
							6 2.2245773629 0.03253336259192303
							7 2.7477258346 0.23853830661339123
							8 48.0860107539 5.2981736126807615
							9 586.6308120187999 153.69744273105638
	        }\nctitime
	        \dotplot{red}{\nctitime};
	        \addplot [name path=upper,draw=none]
	        	table[x=x,y expr=\thisrow{y}+\thisrow{err}] {\nctitime};
	        \addplot [name path=lower,draw=none] 
	        	table[x=x,y expr=\thisrow{y}-\thisrow{err}] {\nctitime};
	        \linedevplot{red}
	        
	    \end{axis}
	    
	\end{tikzpicture}
	\caption{
		A comparison between the number of counterexamples found in the major
		iteration (cti-count) on the left axis and the total runtime (cti-time)
		on the right axis.
	}
	\label{fig:pebbling:relax:5bit:cti}
	\end{subfigure}
	
		\vspace{1em}

	\begin{subfigure}{\textwidth}
		\centering
	\begin{tikzpicture}		
		\begin{axis}
		[
			ymin=0,
			xtick=data,
			xtick style={draw=none},
			minor tick num=1,
			height=4cm,
			width=0.8\textwidth,
			enlarge x limits=0.1,
			enlarge y limits={upper=0},
			ybar,
			axis y line*=right,
			xticklabels={,,},
			bar width=7pt,
			legend style={at={(0.52,1.45)}, anchor=north,legend columns=-1},
			ylabel={Copyrate (\%)},
			fill between/on layer={main},
		]
			\legend{\ipdr copyrate}
			\pgfplotstableread{
					x y err
					7 51.72413793103449 7.105427357601002e-15
					8 76.47058823529413 1.4210854715202004e-14
					9 52.20568488665242 2.1194872906195674
			}\incperc
			\errorbarplot{blue}{\incperc};
		\end{axis}
	
		\begin{axis}
		[
			xtick=data,
			xtick style={draw=none},
			minor tick num=1,
			height=4cm,
			width=0.8\textwidth,
			enlarge x limits=0.05,
			enlarge y limits={upper=0},
			xlabel={Maximum Number of Pebbles},
			axis y line*=left,
			legend style={at={(0.52,1.23)}, anchor=north,legend columns=-1},
			ylabel={Time (s)},
			fill between/on layer={main},
		]
			\legend{\ripdr inc-time, naive inc-time}
			\pgfplotstableread{
				x y err
				7 0.0190888896 0.00012600821834563006
				8 0.028488885700000004 0.0026723702494990875
				9 0.1016615855 0.006221517178199484
			}\inctime
			\dotplot{blue}{\inctime};
			\addplot [name path=upper,draw=none]
				table[x=x,y expr=\thisrow{y}+\thisrow{err}] {\inctime};
			\addplot [name path=lower,draw=none] 
				table[x=x,y expr=\thisrow{y}-\thisrow{err}] {\inctime};
			\filldevplot{blue}
			\pgfplotstableread{
			    x y err
					7 0.0651517197 0.00020768500046996615
					8 0.06513861230000001 0.00025653282489695263
					9 0.06637788279999998 0.0003119165391654623
			}\ninctime
		\dotplot{red}{\ninctime};
		\addplot [name path=upper,draw=none]
			table[x=x,y expr=\thisrow{y}+\thisrow{err}] {\ninctime};
		\addplot [name path=lower,draw=none] 
			table[x=x,y expr=\thisrow{y}-\thisrow{err}] {\ninctime};
		\linedevplot{red}
		\end{axis}
	\end{tikzpicture}
	\caption{
		The amount of the runtime that was spent on copying clauses to the
		relaxed instance denoted on the x-axis. This number is already included
		in the times of the upper graph. 
	}
	\label{fig:pebbling:relax:5bit:inc}
	\end{subfigure}

    \caption{
	    Statistics for the relaxing \texttt{5bitadder} experiment,
	    averaged over 10 repetitions.
    }
    \label{fig:pebbling:relax:5bit}
\end{figure}
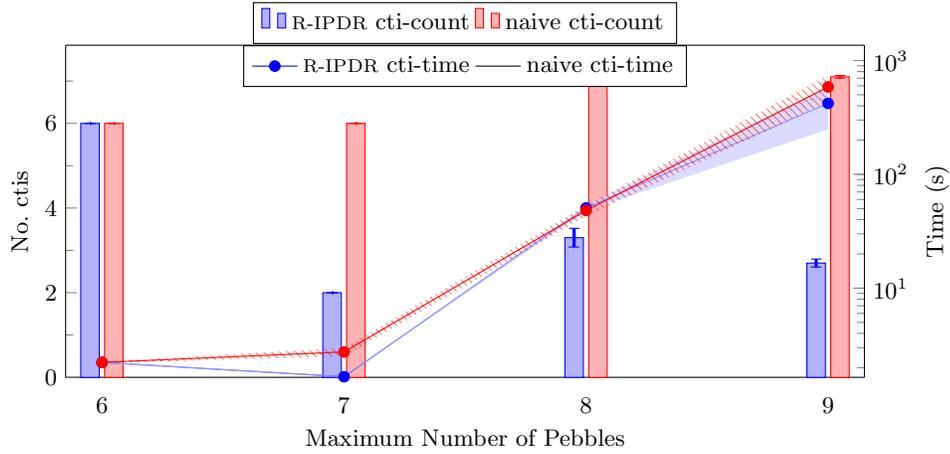
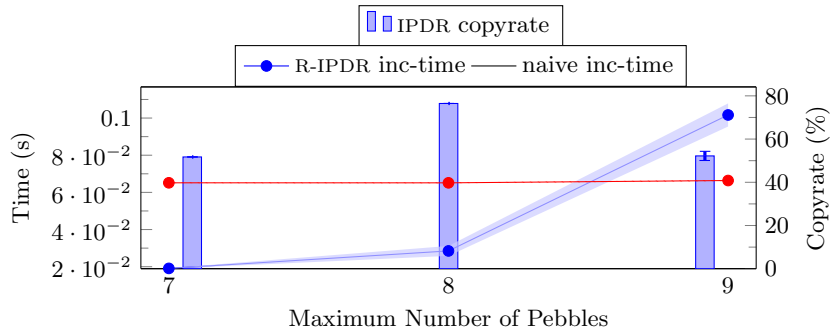

	\paragraph{Relaxing 5bitadder}
	To confirm the previous results on relaxing, we gathered statistics for the \texttt{5bitadder} circuit.
	\autoref{fig:pebbling:relax:5bit} and
	\autoref{fig:pebbling:relax:5bit:oblsat} show the detailed statistics, which did produce 26\% speedup with relaxing \ipdr.
	The number of CTIs in \autoref{fig:pebbling:relax:5bit:cti} and
	obligations handled by \pdr in \autoref{fig:pebbling:relax:5bit:obl} are
	still consistently decreased using \ripdr. 
	 However, this only results in a significant runtime decrease for the last incremental run (9).
	\autoref{fig:pebbling:relax:5bit:sat} shows similar statistics for the number of obligations and \sat calls. 
		Finally, the runtime spent on copying blocking clauses between iterations is negligible as \autoref{fig:pebbling:relax:5bit:inc} shows, while also around 60\% can be copied between runs, as was the case for the \texttt{ham7tc} circuit.
	
\begin{figure}
\centering
\begin{tabular}
{|r|rr|r|}
\hline
 	& \ripdr & Naive \ripdr & Improvement \\\hline
\hline
avg time & 6328.118 s & 10321.160 s & 38.69 \% \\
std dev time & 1788.829 s & 1279.712 s &  \\
avg incremental time & 111.919 s &  &  \\
std dev incremental time & 34.766 s &  &  \\
all hold & yes & yes &  \\
\hline
\end{tabular}

		\vspace{1em}

\begin{subfigure}{\textwidth}
  \begin{tikzpicture}
    \begin{axis}
      [
      xtick=data,
      xtick style={draw=none},
      x axis line style={draw=none},
      minor tick num=1,
      height=6cm,
      width=\textwidth,
      enlarge x limits=0.05,
      enlarge y limits={upper=0},
      xlabel={Maximum Number of Pebbles},
      ybar,
      axis y line*=left,
      bar width=7pt,
      legend style={at={(0.5,1.13)}, anchor=north,legend columns=-1},
      ylabel={No. ctis},
      fill between/on layer={main},
      ]
      \legend{\ripdr cti-count, naive cti-count}
        \pgfplotstableread{
          x y err
		      0 85.33333333333333 15.621684833494658
          1 217.66666666666666 44.67703030682587
          2 580.3333333333334 110.29069333827324
          3 440 49.9799959983992
        }\cticount
        \errorbarplot{blue}{\cticount};
          \pgfplotstableread{
            x y err
            0 85.33333333333333 15.621684833494658
            1 248.33333333333334 33.76607984572903
            2 593.6666666666666 45.25524319940217
            3 1094.3333333333333 21.739194846925535
          }\ncticount
          \errorbarplot{red}{\ncticount};
        \end{axis}
        
        \begin{axis}
          [
          ymode=log,
          log origin=infty,
          xtick=data,
          xtick style={draw=none},
          minor tick num=1,
          height=6cm,
          width=\textwidth,
          enlarge x limits=0.05,
          enlarge y limits={upper=0},
          axis y line*=right,
          legend style={at={(0.5,1.0)}, anchor=north,legend columns=-1},
          xticklabels={,,},
          ylabel={Time (s)},
          fill between/on layer={main},
          ]
          \legend{\ripdr cti-time, naive cti-time}
            \pgfplotstableread{
              x y err
              0 38.503907576333326 9.890703772173731
              1 299.6903125576667 72.81465946629963
              2 2435.83536844 682.6948459075649
              3 3442.169799587 1003.8758792540691
            }\ctitime
            \dotplot{blue}{\ctitime};
            \addplot [name path=upper,draw=none]
            table[x=x,y expr=\thisrow{y}+\thisrow{err}] {\ctitime};
            \addplot [name path=lower,draw=none] 
            table[x=x,y expr=\thisrow{y}-\thisrow{err}] {\ctitime};
            \filldevplot{blue}

              \pgfplotstableread{
                x y err
                0 38.603902608333335 9.89382603041477
                1 273.68145131533333 39.25475183590951
                2 1339.555998946 263.79998686654864
                3 8667.869587788999 996.0010019415879
              }\nctitime
              \dotplot{red}{\nctitime};
              \addplot [name path=upper,draw=none]
              table[x=x,y expr=\thisrow{y}+\thisrow{err}] {\nctitime};
              \addplot [name path=lower,draw=none] 
              table[x=x,y expr=\thisrow{y}-\thisrow{err}] {\nctitime};
              \linedevplot{red}
              
            \end{axis}
            
          \end{tikzpicture}
          \caption{
            A comparison between the number of counterexamples found in the major
            iteration (cti-count) on the left axis and the total runtime (cti-time)
            on the right axis.
          }
          \label{fig:peter:relax:proc4:cti}
\end{subfigure}

		\vspace{1em}

\begin{subfigure}{\textwidth}
  \centering
  \begin{tikzpicture}
    \begin{axis}
    [
    ymin=0,
    xtick=data,
    xtick style={draw=none},
    minor tick num=1,
    height=4cm,
    width=0.8\textwidth,
    enlarge x limits=0.1,
    enlarge y limits={upper=0},
    ybar,
    axis y line*=right,
    xticklabels={,,},
    bar width=7pt,
    legend style={at={(0.52,1.45)}, anchor=north,legend columns=-1},
    ylabel={Copyrate (\%)},
    fill between/on layer={main},
    ]
    \legend{\ripdr copyrate}
      \pgfplotstableread{
        x y err
        1 34.38237952819207 4.507638444841585
        2 43.93369297918042 3.8094062104834636
        3 42.99150156629475 6.951986358854146
      }\incperc
      \errorbarplot{blue}{\incperc};
    \end{axis}
    
    \begin{axis}
      [
      ymode=log,
      log origin=infty,
      xtick=data,
      xtick style={draw=none},
      minor tick num=1,
      height=4cm,
      width=0.8\textwidth,
      enlarge x limits=0.1,
      enlarge y limits={upper=0},
      xlabel={Maximum number of context switches},
      axis y line*=left,
      legend style={at={(0.5,1.1)}, anchor=north,legend columns=-1},
      ylabel={Time (s)},
      ]
      \legend{\ripdr inc-time, naive inc-time}
        \pgfplotstableread{
          x y err
          1 1.591105994 0.3820612356361238
          2 9.134560197333332 1.3397921296984834
          3 101.19338122900001 33.42184929164526
        }\inctime
        \dotplot{blue}{\inctime};
        \addplot [name path=upper,draw=none]
        table[x=x,y expr=\thisrow{y}+\thisrow{err}] {\inctime};
        \addplot [name path=lower,draw=none] 
        table[x=x,y expr=\thisrow{y}-\thisrow{err}] {\inctime};
        \filldevplot{blue}
          \pgfplotstableread{
            x y err
            1 0.47466611966666666 0.0009166446212170848
            2 0.486427065 0.0031833461686044525
            3 0.4882934906666667 0.005215346325370431
          }\ninctime
          \dotplot{red}{\ninctime};
          \addplot [name path=upper,draw=none]
          table[x=x,y expr=\thisrow{y}+\thisrow{err}] {\ninctime};
          \addplot [name path=lower,draw=none] 
          table[x=x,y expr=\thisrow{y}-\thisrow{err}] {\ninctime};
          \linedevplot{red}
        \end{axis}
      \end{tikzpicture}
      \caption{
        The amount of the runtime that was spent on copying clauses to the
        relaxed instance denoted on the x-axis. This number is already included
        in the times of the upper graph. The bars denote the percentage (right y-axis) of clauses that could be copied to the instance with the bound on the x-axis.
      }\label{fig:peter:relax:4proc:inc}
  \end{subfigure}
  \caption{
    Statistics for the relaxing \ttfunction{Peterson4} experiment,
    averaged over 3 repetitions.
  }
  \label{fig:peter:relax:4proc}
  
\end{figure}
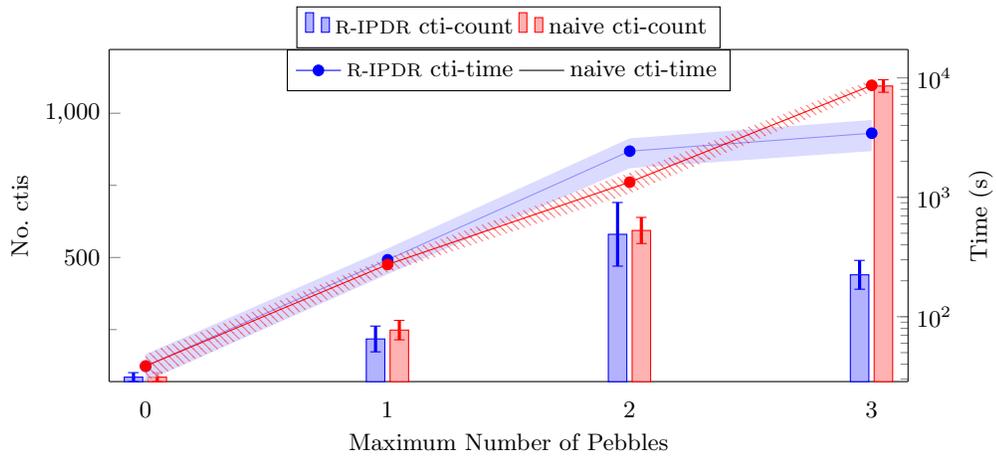
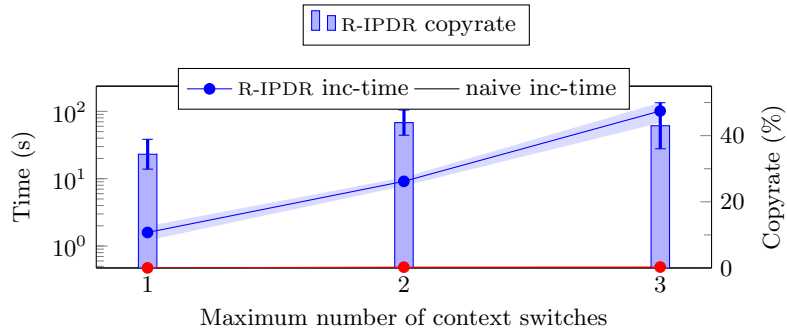

	\paragraph{Relaxing Peterson4}
	The relaxing \ipdr run on the \texttt{Peterson4} protocol shows a similar behavior as the 
	\texttt{5bitadder} discussed above (see \autoref{fig:peter:relax:4proc} and \autoref{fig:peter:relax:4proc:oblsat}). While only 40\% of the clauses can be copied accross incremental runs, the runtime benefits in the final increment are more pronounced.

{
\newcommand{\plotheight}{6cm}
	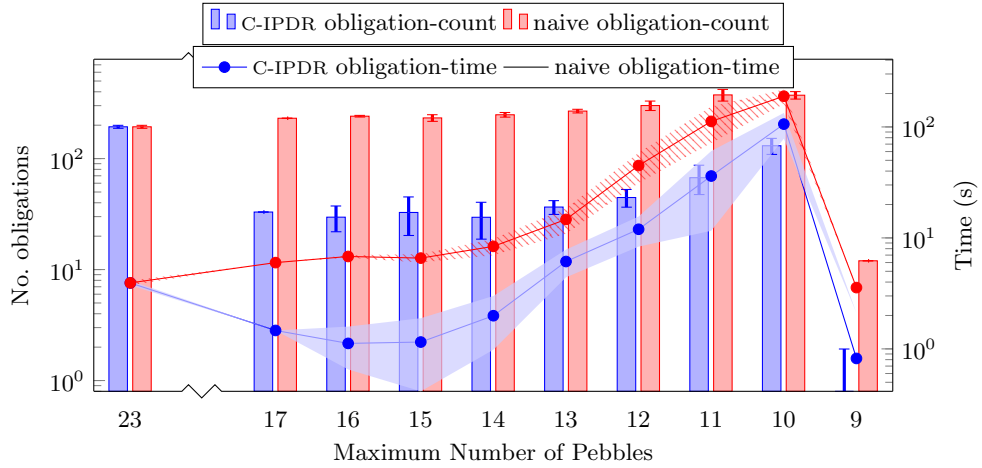
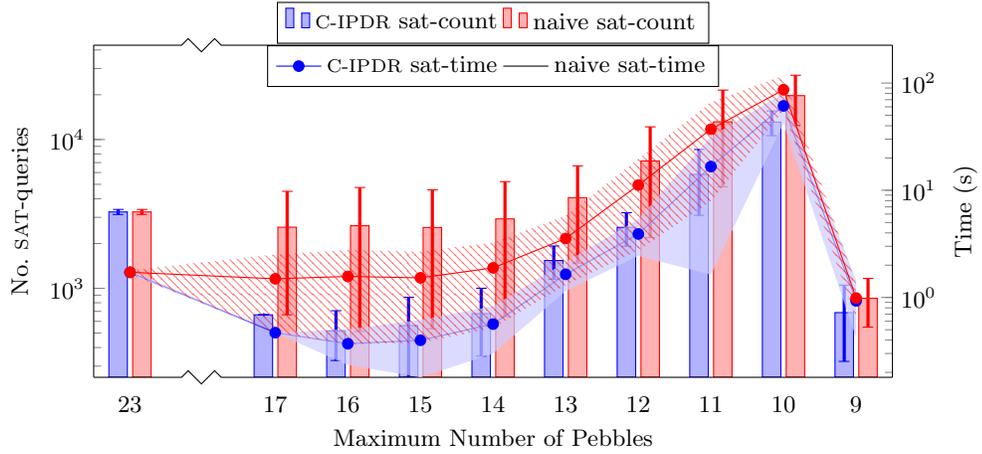
\begin{figure}[h]
	\centering
	\begin{subfigure}{\textwidth}
	\begin{tikzpicture}
	    \begin{axis}
		[
        	ymode=log,
			log origin=infty,
			xtick=data,
			xtick style={draw=none},
			x axis line style={draw=none},
			minor tick num=1,
			height=\plotheight,
			width=\textwidth,
			enlarge x limits=0.05,,
			enlarge y limits={upper=0},
			xlabel={Maximum Number of Pebbles},
			xticklabels={9,10,11,12,13,14,15,16,17,23},
			ybar,
			axis y line*=left,
			bar width=7pt,
			legend style={at={(0.5,1.17)}, anchor=north,legend columns=-1},
			ylabel={No. obligations},
			x dir = reverse,
			fill between/on layer={main},
		]		
	        \legend{\cipdr obligation-count, naive obligation-count}
	        \pgfplotstableread{
	        	x y err
	            9 0.8 1.1242775458044156
	            10 130.6 21.567475512910637
	            11 67.3 19.87785199663183
	            12 44.6 8.10530690350464
	            13 36.6 5.199615370390391
	            14 29.6 10.799814813227124
	            15 32.714285714285715 12.426577367597496
	            16 29.666666666666668 7.748357054049396
	            17 33 0
	            19 193.5 6.002082971769051
	        }\oblcount
        	\errorbarplot{blue}{\oblcount};
			\pgfplotstableread{
	            x y err
	            9 12 0
	            10 372.6 28.83289787725126
	            11 375 45.21061822182926
	            12 300.6 29.362833650722475
	            13 268 10.16857905510893
	            14 248.3 11.933524207039595
	            15 232.57142857142858 16.047306159770955
	            16 241 4.08248290463863
	            17 231 0
	            19 193.5 6.002082971769051
	        }\noblcount
        	\errorbarplot{red}{\noblcount};
	        
      		\defxcrunch{0.11}{0.16}
      	\end{axis}
      
		\drawxcrunch

	    \begin{axis}
        [
			ymode=log,
			log origin=infty,
			xtick=data,
			xtick style={draw=none},
			x axis line style={draw=none},
			minor tick num=1,
			height=\plotheight,
			width=\textwidth,
			enlarge x limits=0.05,,
			enlarge y limits={upper=0},
			axis y line*=right,
			legend style={at={(0.5,1.04)}, anchor=north,legend columns=-1},
			xticklabels={,,},
			ylabel={Time (s)},
			x dir = reverse,
			fill between/on layer={main},
		]
	        \legend{\cipdr obligation-time, naive obligation-time}
			\pgfplotstableread{
	            x y err
	            9 0.8202922705999999 1.2885606330432002
	            10 106.0217300456 27.66452546685597
	            11 36.1087847796 24.522193301870765
	            12 11.9647086355 3.6186424296488027
	            13 6.1279803073000005 1.7154989758489123
	            14 1.9904427260000002 1.008286755695432
	            15 1.1523403985714284 0.7376926638007634
	            16 1.120595703333333 0.4678302175091482
	            17 1.4657357190000002 0
	            19 3.9288569940000015 0.19449981237201608
	        }\obltime
	        \dotplot{blue}{\obltime};
		    \addplot [name path=upper,draw=none]
				table[x=x,y expr=\thisrow{y}+\thisrow{err}] {\obltime};
			\addplot [name path=lower,draw=none] 
				table[x=x,y expr=\thisrow{y}-\thisrow{err}] {\obltime};
			\filldevplot{blue}
			\pgfplotstableread{
	            x y err
	            9 3.5680185307000003 0.01813684335066361
	            10 189.29807288689992 28.54660661977086
	            11 112.19739899669999 40.50656380232495
	            12 44.868530388900005 13.668511095333836
	            13 14.682385296499998 3.8080319991172944
	            14 8.390014096000002 1.4128448844761268
	            15 6.577928030428572 0.5814390788171226
	            16 6.803558257333333 0.1774575614997469
	            17 5.989378324 0
	            19 3.9347875008999997 0.1952722059439133
	        }\nobltime
        \dotplot{red}{\nobltime};
	    \addplot [name path=upper,draw=none]
			table[x=x,y expr=\thisrow{y}+\thisrow{err}] {\nobltime};
		\addplot [name path=lower,draw=none] 
			table[x=x,y expr=\thisrow{y}-\thisrow{err}] {\nobltime};
		\linedevplot{red}
	    \end{axis}
	    
	\end{tikzpicture}
	\caption{
	    Left axis: total number of obligations (obligation-count)
	    handled by the minor iteration. Right axis: the time spent on
	    them (obligation-time).
	}\label{fig:pebbling:constrain:ham7:obl}
	\end{subfigure}
	
		\vspace{1em}
		
	\begin{subfigure}{\textwidth}
	\begin{tikzpicture}
	    \begin{axis}
        [
			ymode=log,
			log origin=infty,
			xtick=data,
			xtick style={draw=none},
			x axis line style={draw=none},
			minor tick num=1,
			height=\plotheight,
			width=\textwidth,
			enlarge x limits=0.05,,
			enlarge y limits={upper=0},
			xlabel={Maximum Number of Pebbles},
			xticklabels={9,10,11,12,13,14,15,16,17,23},
			ybar,
			axis y line*=left,
			bar width=7pt,
			legend style={at={(0.5,1.13)}, anchor=north,legend columns=-1},
			ylabel={No. \sat-queries},
			x dir = reverse,
			fill between/on layer={main},
		]
	        \legend{\cipdr sat-count, naive sat-count}
        	\pgfplotstableread{
	            x y err
	            9 686.8 364.0208840162883
	            10 13094.6 2475.66733952686
	            11 5834 2739.3741621034537
	            12 2569.8 653.9715161992914
	            13 1538.3 387.40370287337214
	            14 675.8 325.6146249786702
	            15 562 309.08343764843085
	            16 517.3333333333334 190.24350867399548
	            17 664 0
	            19 3261.9 128.06990669161902
	        }\satcount
        	\errorbarplot{blue}{\satcount}
			\pgfplotstableread{
	            x y err
	            9 857.6 308.92510095490786
	            10 19733.9 7310.827689837861
	            11 13160 8348.426013327304
	            12 7180.6 4989.815138659948
	            13 4066.9 2588.1654758728237
	            14 2929.3 2284.5051662230926
	            15 2564.214285714286 2028.9478607512885
	            16 2635.1666666666665 2122.8429193384013
	            17 2575.5 1911.4999673031648
	            19 3261.9 128.06990473956012
			}\nsatcount
			\errorbarplot{red}{\nsatcount}
	        
			\defxcrunch{0.11}{0.16}
		\end{axis}
	
		\drawxcrunch
	
	    \begin{axis}
		[
			ymode=log,
			log origin=infty,
			xtick=data,
			xtick style={draw=none},
			x axis line style={draw=none},
			minor tick num=1,
			height=\plotheight,
			width=\textwidth,
			enlarge x limits=0.05,,
			enlarge y limits={upper=0},
			axis y line*=right,
			legend style={at={(0.5,1.0)}, anchor=north,legend columns=-1},
			xticklabels={,,},
			ylabel={Time (s)},
			x dir = reverse,
			fill between/on layer={main},
		]
	        \legend{\cipdr sat-time, naive sat-time}
        	\pgfplotstableread{
	            x y err
	            9 0.9294551663000004 0.5096581628267625
	            10 61.24580778559999 19.104282959821106
	            11 16.635235819399995 15.00663173900241
	            12 3.911259109600002 1.4701238475392202
	            13 1.6449584765 0.49634482217972353
	            14 0.5641894206000002 0.26486389230701185
	            15 0.398716786142857 0.21887224695394272
	            16 0.3700911879999998 0.13554958154612798
	            17 0.4691777069999996 0
	            19 1.7160446387000008 0.07834621710623756
	        }\sattime
	        \dotplot{blue}{\sattime};
		    \addplot [name path=upper,draw=none]
				table[x=x,y expr=\thisrow{y}+\thisrow{err}] {\sattime};
			\addplot [name path=lower,draw=none] 
				table[x=x,y expr=\thisrow{y}-\thisrow{err}] {\sattime};
			\filldevplot{blue}
			\pgfplotstableread{
	            x y err
	            9 0.9875451029500006 0.365121869178411
	            10 86.60907545644997 31.26317670300262
	            11 37.106986784400014 29.236260238603933
	            12 11.17557922265001 8.43385292259983
	            13 3.5501318568499984 2.2473711978595503
	            14 1.8885429831000002 1.370453120541001
	            15 1.5246994530000002 1.1480536092761238
	            16 1.5704203605 1.2047731275108726
	            17 1.4935606224999978 1.0243829154999982
	            19 1.717350948450001 0.07858515107129796
	        }\nsattime
        \dotplot{red}{\nsattime};
	    \addplot [name path=upper,draw=none]
			table[x=x,y expr=\thisrow{y}+\thisrow{err}] {\nsattime};
		\addplot [name path=lower,draw=none] 
			table[x=x,y expr=\thisrow{y}-\thisrow{err}] {\nsattime};
		\linedevplot{red}
	        
	    \end{axis}
	\end{tikzpicture}
	\caption{
	    Left axis: the total number of calls to the \sat-solver (sat-count). Right axis: the total time spent waiting on the \sat-solver (sat-time..
	}\label{fig:pebbling:constrain:ham7:sat}
	\end{subfigure}
	\caption{
		Constraining \texttt{ham7tc} experiment (10 repetitions). Left axis: the number of counterexamples (cti-count) found in the major iteration. Right axis: the total runtime (cti-time).
		On constraint 23, the algorithm found a trace with 18 pebbles, so it could continue with 17 afterwards.
	}
	\label{fig:pebbling:constrain:ham7:oblsat}
	\end{figure}
}

{
\newcommand{\plotheight}{6cm}
	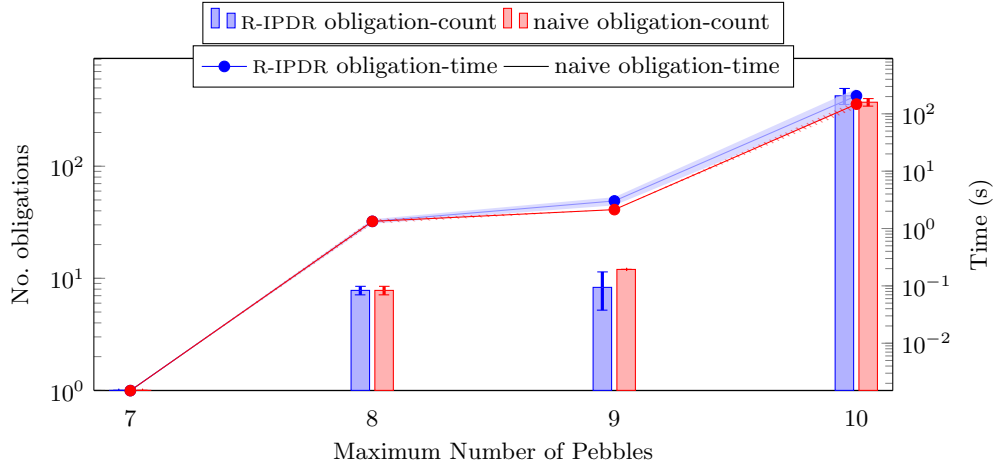
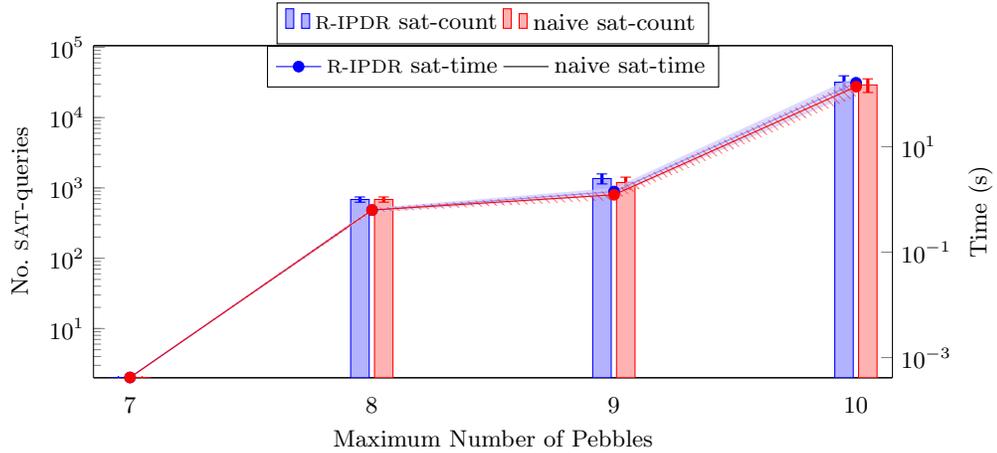
\begin{figure}[h]
	\centering
	\begin{subfigure}{\textwidth}
	\begin{tikzpicture}
	    \begin{axis}
			[
				 ymode=log,
				 log origin=infty,
				xtick=data,
				xtick style={draw=none},
				minor tick num=1,
				height=\plotheight,
				width=\textwidth,
				enlarge x limits=0.05,,
				enlarge y limits={upper=0},
				xlabel={Maximum Number of Pebbles},
				ybar,
				axis y line*=left,
				bar width=7pt,
				legend style={at={(0.5,1.17)}, anchor=north,legend columns=-1},
				ylabel={No. obligations},
				fill between/on layer={main},
			]		
	        \legend{\ripdr obligation-count, naive obligation-count}
	        \pgfplotstableread{
							x y err
							7 1 0
							8 7.8 0.681175454637056
							9 8.3 3.103063002905355
							10 424.9 69.22774732720978
	        }\oblcount
        	\errorbarplot{blue}{\oblcount};
					\pgfplotstableread{
							x y err
							7 1 0
							8 7.8 0.681175454637056
							9 12 0
							10 372.6 28.83289787725126
	        }\noblcount
        	\errorbarplot{red}{\noblcount};
      	\end{axis}

	    \begin{axis}
        [
          ymode=log,
          log origin=infty,
					xtick=data,
					xtick style={draw=none},
					minor tick num=1,
					height=\plotheight,
					width=\textwidth,
					enlarge x limits=0.05,,
					enlarge y limits={upper=0},
					axis y line*=right,
					legend style={at={(0.5,1.04)}, anchor=north,legend columns=-1},
					xticklabels={,,},
					ylabel={Time (s)},
					fill between/on layer={main},
				]
	        \legend{\ripdr obligation-time, naive obligation-time}
					\pgfplotstableread{
	            x y err
							7 0.0014916292999999998 0
							8 1.3327998512 0.12149405186689703
							9 3.0275430550999993 0.4908335124356978
							10 206.24312188169998 67.61955351235518
	        }\obltime
	        \dotplot{blue}{\obltime};
					\addplot [name path=upper,draw=none]
						table[x=x,y expr=\thisrow{y}+\thisrow{err}] {\obltime};
					\addplot [name path=lower,draw=none] 
						table[x=x,y expr=\thisrow{y}-\thisrow{err}] {\obltime};
					\filldevplot{blue}
					\pgfplotstableread{
	            x y err
							7 0.0014905496 0
							8 1.3328906591999998 0.12247734019036084
							9 2.1362515233999995 0.013411346549398682
							10 147.58029648619998 22.423577909455354
	        }\nobltime
        \dotplot{red}{\nobltime};
	    \addplot [name path=upper,draw=none]
			table[x=x,y expr=\thisrow{y}+\thisrow{err}] {\nobltime};
		\addplot [name path=lower,draw=none] 
			table[x=x,y expr=\thisrow{y}-\thisrow{err}] {\nobltime};
		\linedevplot{red}
	    \end{axis}
	    
	\end{tikzpicture}
	\caption{
	    Left axis: total number of obligations (obligation-count)
	    handled by the minor iteration. Right axis: the time spent on
	    them (obligation-time). For constraint 7, no obligations were enqueued as \pdr terminated in the initialization step. Its obligation-count value was set to 1 to accommodate the logarithmic scale and the obligation-time to the total runtime (which is a close approximation).
	}\label{fig:pebbling:relax:ham7:obl}
	\end{subfigure}

		\vspace{1em}
	
	\begin{subfigure}{\textwidth}
	\begin{tikzpicture}
	    \begin{axis}
        [
					ymode=log,
					log origin=infty,
					xtick=data,
					xtick style={draw=none},
					minor tick num=1,
					height=\plotheight,
					width=\textwidth,
					enlarge x limits=0.05,,
					enlarge y limits={upper=0},
					xlabel={Maximum Number of Pebbles},
					ybar,
					axis y line*=left,
					bar width=7pt,
					legend style={at={(0.5,1.13)}, anchor=north,legend columns=-1},
					ylabel={No. \sat-queries},
					fill between/on layer={main},
				]
	        \legend{\ripdr sat-count, naive sat-count}
        	\pgfplotstableread{
	            x y err
							7 2 0
							8 684.8 58.544376331121676
							9 1359.5 220.39515648035461
							10 31871.3 7394.4379643756565
	        }\satcount
        	\errorbarplot{blue}{\satcount}
					\pgfplotstableread{
	            x y err
							7 2 0
							8 685.3 58.54651569478751
							9 1193.95 227.3769230265024
							10 29122.25 6418.688727312223
					}\nsatcount
					\errorbarplot{red}{\nsatcount}
			\end{axis}
	
	    \begin{axis}
		[
			ymode=log,
			log origin=infty,
			xtick=data,
			xtick style={draw=none},
			minor tick num=1,
			height=\plotheight,
			width=\textwidth,
			enlarge x limits=0.05,,
			enlarge y limits={upper=0},
			axis y line*=right,
			legend style={at={(0.5,1.0)}, anchor=north,legend columns=-1},
			xticklabels={,,},
			ylabel={Time (s)},
			fill between/on layer={main},
		]
	        \legend{\ripdr sat-time, naive sat-time}
        	\pgfplotstableread{
	            x y err
							7 0.0004199052 6.501300774460455e-06
							8 0.6288624799999999 0.05748040813507536
							9 1.411619574 0.2382303543920528
							10 163.16516252129983 57.13601883085018
	        }\sattime
	        \dotplot{blue}{\sattime};
					\addplot [name path=upper,draw=none]
						table[x=x,y expr=\thisrow{y}+\thisrow{err}] {\sattime};
					\addplot [name path=lower,draw=none] 
						table[x=x,y expr=\thisrow{y}-\thisrow{err}] {\sattime};
					\filldevplot{blue}
					\pgfplotstableread{
	            x y err
							7 0.00041880650000000004 4.804880492790643e-06
							8 0.6291450257 0.05759535313548981
							9 1.2262487040000003 0.2506104665708615
							10 137.97937861644988 49.2045161793266
	        }\nsattime
        \dotplot{red}{\nsattime};
				\addplot [name path=upper,draw=none]
					table[x=x,y expr=\thisrow{y}+\thisrow{err}] {\nsattime};
				\addplot [name path=lower,draw=none] 
					table[x=x,y expr=\thisrow{y}-\thisrow{err}] {\nsattime};
				\linedevplot{red}
	    \end{axis}
	\end{tikzpicture}
	\caption{
			Left axis: the total number of calls to the \sat-solver (sat-count).
			Right axis: the total time spent waiting on the \sat-solver (sat-time..
	}\label{fig:pebbling:relax:ham7:sat}
	\end{subfigure}
	\caption{
		Relaxing \texttt{ham7tc} experiment (10 repetitions). Left axis: the number
		of counterexamples (cti-count) found in the major iteration. Right axis:
		the total runtime (cti-time).
	}
	\label{fig:pebbling:relax:ham7:oblsat}
	\end{figure}
}

{
\newcommand{\plotheight}{6cm}
	\begin{figure}[h]
	\centering
	\begin{subfigure}{\textwidth}
	\begin{tikzpicture}
	    \begin{axis}
			[
				ymode=log,
				log origin=infty,
				xtick=data,
				xtick style={draw=none},
				minor tick num=1,
				height=\plotheight,
				width=\textwidth,
				enlarge x limits=0.05,,
				enlarge y limits={upper=0},
				xlabel={Maximum Number of Pebbles},
				ybar,
				axis y line*=left,
				bar width=7pt,
				legend style={at={(0.5,1.17)}, anchor=north,legend columns=-1},
				ylabel={No. obligations},
				fill between/on layer={main},
			]		
	        \legend{\ripdr obligation-count, naive obligation-count}
	        \pgfplotstableread{
							x y err
							6 14 0
							7 5 0
							8 90.6 3.3490297102295172
							9 398.2 61.63265368293012
	        }\oblcount
        	\errorbarplot{blue}{\oblcount};
					\pgfplotstableread{
							x y err
							6 14 0
							7 17.5 1.4230249470757708
							8 99.3 5.738379562210921
							9 454.7 51.886115676546844
	        }\noblcount
        	\errorbarplot{red}{\noblcount};
      	\end{axis}

	    \begin{axis}
			[
				ymode=log,
				log origin=infty,
				xtick=data,
				xtick style={draw=none},
				minor tick num=1,
				height=\plotheight,
				width=\textwidth,
				enlarge x limits=0.05,,
				enlarge y limits={upper=0},
				axis y line*=right,
				legend style={at={(0.5,1.04)}, anchor=north,legend columns=-1},
				xticklabels={,,},
				ylabel={Time (s)},
				fill between/on layer={main},
			]
	        \legend{\ripdr obligation-time, naive obligation-time}
					\pgfplotstableread{
	            x y err
							6 2.1623089361999996 0.03232719238938985
							7 1.6105662851000002 0.019689068461510296
							8 50.30251861420001 2.879688859163332
							9 419.2108461142 171.80678459713639
	        }\obltime
	        \dotplot{blue}{\obltime};
					\addplot [name path=upper,draw=none]
						table[x=x,y expr=\thisrow{y}+\thisrow{err}] {\obltime};
					\addplot [name path=lower,draw=none] 
						table[x=x,y expr=\thisrow{y}-\thisrow{err}] {\obltime};
					\filldevplot{blue}
					\pgfplotstableread{
	            x y err
							6 2.1647121930999997 0.0322873266842592
							7 2.6820871597 0.2335312158298106
							8 47.683364728799994 5.275170589545306
							9 585.9060500269001 153.68933684833615
	        }\nobltime
        \dotplot{red}{\nobltime};
	    \addplot [name path=upper,draw=none]
			table[x=x,y expr=\thisrow{y}+\thisrow{err}] {\nobltime};
		\addplot [name path=lower,draw=none] 
			table[x=x,y expr=\thisrow{y}-\thisrow{err}] {\nobltime};
		\linedevplot{red}
	    \end{axis}
	    
	\end{tikzpicture}
	\caption{
	    Left axis: total number of obligations (obligation-count)
	    handled by the minor iteration. Right axis: the time spent on
	    them (obligation-time).
	}\label{fig:pebbling:relax:5bit:obl}
	\end{subfigure}
	
		\vspace{1em}
	
	\begin{subfigure}{\textwidth}
	\begin{tikzpicture}
	    \begin{axis}
			[
				ymode=log,
				log origin=infty,
				xtick=data,
				xtick style={draw=none},
				minor tick num=1,
				height=\plotheight,
				width=\textwidth,
				enlarge x limits=0.05,,
				enlarge y limits={upper=0},
				xlabel={Maximum Number of Pebbles},
				ybar,
				axis y line*=left,
				bar width=7pt,
				legend style={at={(0.5,1.13)}, anchor=north,legend columns=-1},
				ylabel={No. \sat-queries},
				fill between/on layer={main},
			]
	        \legend{\ripdr sat-count, naive sat-count}
        	\pgfplotstableread{
	            x y err
							6 1051.5 10.120523701864444
							7 734.5 3.664014192112252
							8 12393.9 556.2368209674726
							9 44375 8556.661498505127
	        }\satcount
        	\errorbarplot{blue}{\satcount}
					\pgfplotstableread{
	            x y err
							6 1051.5 10.119906126046821
							7 1023.85 296.4442715334537
							8 12332.85 796.6394109790201
							9 46827.85 8085.4436595109
					}\nsatcount
					\errorbarplot{red}{\nsatcount}
			\end{axis}
	
	    \begin{axis}
			[
				ymode=log,
				log origin=infty,
				xtick=data,
				xtick style={draw=none},
				minor tick num=1,
				height=\plotheight,
				width=\textwidth,
				enlarge x limits=0.05,,
				enlarge y limits={upper=0},
				axis y line*=right,
				legend style={at={(0.5,1.0)}, anchor=north,legend columns=-1},
				xticklabels={,,},
				ylabel={Time (s)},
				fill between/on layer={main},
			]
	        \legend{\ripdr sat-time, naive sat-time}
        	\pgfplotstableread{
	            x y err
							6 0.9481353569 0.019532439270550823
							7 0.7177197675999999 0.013498250778513226
							8 31.763181837400026 2.5212951111906072
							9 344.54804534880043 154.3420728523468
	        }\sattime
	        \dotplot{blue}{\sattime};
					\addplot [name path=upper,draw=none]
						table[x=x,y expr=\thisrow{y}+\thisrow{err}] {\sattime};
					\addplot [name path=lower,draw=none] 
						table[x=x,y expr=\thisrow{y}-\thisrow{err}] {\sattime};
					\filldevplot{blue}
					\pgfplotstableread{
	            x y err
							6 0.9487566032999994 0.019615630891547196
							7 0.9546464854500003 0.24824697272922058
							8 30.859400666250014 3.3531731894325594
							9 422.8443872479 166.72516507039344
	        }\nsattime
        \dotplot{red}{\nsattime};
				\addplot [name path=upper,draw=none]
					table[x=x,y expr=\thisrow{y}+\thisrow{err}] {\nsattime};
				\addplot [name path=lower,draw=none] 
					table[x=x,y expr=\thisrow{y}-\thisrow{err}] {\nsattime};
				\linedevplot{red}
	    \end{axis}
	\end{tikzpicture}
	\caption{
			Left axis: the total number of calls to the \sat-solver (sat-count).
			Right axis: the total time spent waiting on the \sat-solver (sat-time..
	}\label{fig:pebbling:relax:5bit:sat}
	\end{subfigure}
	\caption{
		Relaxing \texttt{5bitadder} experiment (10 repetitions). Left axis: the number
		of counterexamples (cti-count) found in the major iteration. Right axis:
		the total runtime (cti-time).
	}
	\label{fig:pebbling:relax:5bit:oblsat}
	\end{figure}
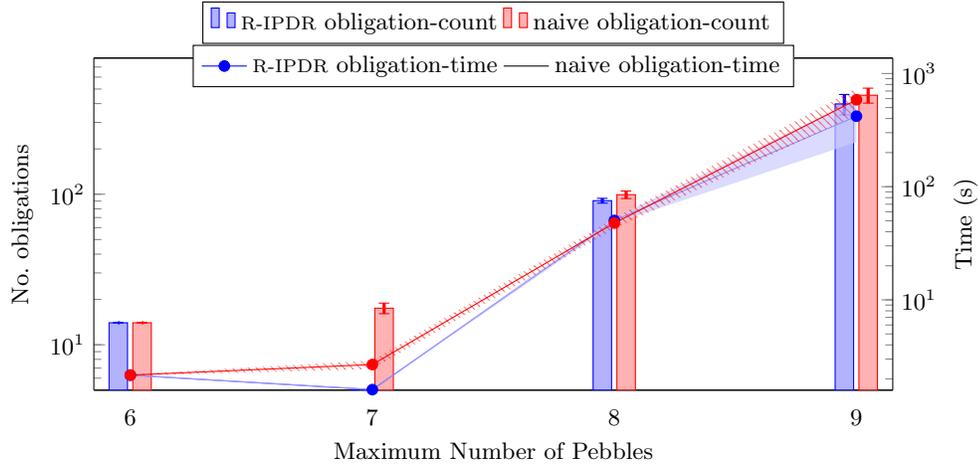
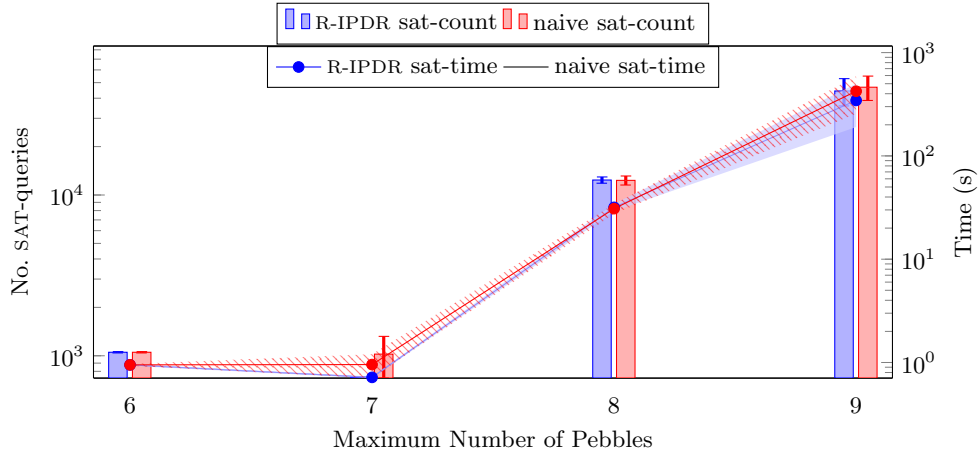
}

{
\newcommand{\plotheight}{6cm}
\begin{figure}[p]
\centering
\vspace*{-2cm}
\begin{subfigure}{\textwidth}
  \begin{tikzpicture}
  \begin{axis}
    [
    ymode=log,
    log origin=infty,
    xtick=data,
    xtick style={draw=none},
    minor tick num=1,
    height=\plotheight,
    width=\textwidth,
    enlarge x limits=0.05,,
    enlarge y limits={upper=0},
    xlabel={Maximum Number of context switches},
    ybar,
    axis y line*=left,
    bar width=7pt,
    legend style={at={(0.5,1.17)}, anchor=north,legend columns=-1},
    ylabel={No. obligations},
    fill between/on layer={main},
    ]		
    \legend{\ripdr obligation-count, naive obligation-count}
    \pgfplotstableread{
      x y err
      0 276.6666666666667 91.08258360980456
      1 578 133.09645625134678
      2 1698 277.6040345528141
      3 1772 280.871263511714
    }\oblcount
    \errorbarplot{blue}{\oblcount};
      \pgfplotstableread{
        x y err
        0 276.6666666666667 91.08258360980456
        1 629 89.54328562209453
        2 1553.6666666666667 190.5702942145943
        3 3936 128.90565025113006
      }\noblcount
      \errorbarplot{red}{\noblcount};
  \end{axis}
  
  \begin{axis}
    [
    ymode=log,
    log origin=infty,
    xtick=data,
    xtick style={draw=none},
    minor tick num=1,
    height=\plotheight,
    width=\textwidth,
    enlarge x limits=0.05,,
    enlarge y limits={upper=0},
    axis y line*=right,
    legend style={at={(0.5,1.04)}, anchor=north,legend columns=-1},
    xticklabels={,,},
    ylabel={Time (s)},
    fill between/on layer={main},
    ]
    \legend{\ripdr obligation-time, naive obligation-time}
      \pgfplotstableread{
        x y err
        0 30.904991565 8.40386617189863
        1 261.71103614066664 65.19025482113047
        2 2211.555925235 620.814298109431
        3 2999.8427126716674 867.1718609971315
      }\obltime
      \dotplot{blue}{\obltime};
      \addplot [name path=upper,draw=none]
      table[x=x,y expr=\thisrow{y}+\thisrow{err}] {\obltime};
      \addplot [name path=lower,draw=none] 
      table[x=x,y expr=\thisrow{y}-\thisrow{err}] {\obltime};
      \filldevplot{blue}
        \pgfplotstableread{
          x y err
          0 31.000653090999993 8.401838004882366
          1 241.47052277166662 36.21597796421707
          2 1223.3217029119996 248.8063503849408
          3 7985.181220806674 932.7863237848143
        }\nobltime
        \dotplot{red}{\nobltime};
        \addplot [name path=upper,draw=none]
        table[x=x,y expr=\thisrow{y}+\thisrow{err}] {\nobltime};
        \addplot [name path=lower,draw=none] 
        table[x=x,y expr=\thisrow{y}-\thisrow{err}] {\nobltime};
        \linedevplot{red}
      \end{axis}
      
    \end{tikzpicture}
    
    \caption{
      This graph compares the total number of obligations (obligation-count)
      handled by the minor iteration on the left axis and the time spent on
      them (obligation-time) on the right axis.
    }\label{fig:peter:relax:4proc:obl}
\end{subfigure}

		\vspace{1em}

\begin{subfigure}{\textwidth}
\begin{tikzpicture}
  \begin{axis}
    [
    ymode=log,
    log origin=infty,
    xtick=data,
    xtick style={draw=none},
    minor tick num=1,
    height=\plotheight,
    width=\textwidth,
    enlarge x limits=0.05,,
    enlarge y limits={upper=0},
    xlabel={Maximum Number of Pebbles},
    ybar,
    axis y line*=left,
    bar width=7pt,
    legend style={at={(0.5,1.13)}, anchor=north,legend columns=-1},
    ylabel={No. \sat-queries},
    fill between/on layer={main},
    ]
    \legend{\ripdr sat-count, naive sat-count}
      \pgfplotstableread{
        x y err
        0 10535 2733.851495601032
        1 29088.666666666668 6347.638496439559
        2 104002.66666666667 14702.164403520594
        3 135698.33333333334 21867.639590482988
      }\satcount
      \errorbarplot{blue}{\satcount}
      \pgfplotstableread{
        x y err
        0 10535 2733.851495601032
        1 29509.833333333332 5461.454736419864
        2 92527.5 17462.612482338754
        3 194242.83333333334 31165.13358714558
      }\nsatcount
      \errorbarplot{red}{\nsatcount}
    \end{axis}
      
    \begin{axis}
      [
      ymode=log,
      log origin=infty,
      xtick=data,
      xtick style={draw=none},
      minor tick num=1,
      height=\plotheight,
      width=\textwidth,
      enlarge x limits=0.05,,
      enlarge y limits={upper=0},
      axis y line*=right,
      legend style={at={(0.5,1.0)}, anchor=north,legend columns=-1},
      xticklabels={,,},
      ylabel={Time (s)},
      fill between/on layer={main},
      ]
      \legend{\ripdr sat-time, naive sat-time}
      \pgfplotstableread{
        x y err
        0 29.412513860666706 7.445968150057138
        1 234.1678813983335 56.67486168227062
        2 2102.0730012673444 619.0070475154997
        3 3095.5007721913607 947.3423316981982
      }\sattime
      \dotplot{blue}{\sattime};
      \addplot [name path=upper,draw=none]
      table[x=x,y expr=\thisrow{y}+\thisrow{err}] {\sattime};
      \addplot [name path=lower,draw=none] 
      table[x=x,y expr=\thisrow{y}-\thisrow{err}] {\sattime};
      \filldevplot{blue}
      \pgfplotstableread{
        x y err
        0 29.44824518783334 7.446754046766824
        1 221.10678567183334 47.329792403208295
        2 1601.6451422278385 684.540733565861
        3 5384.918382473014 2469.8700492843986
      }\nsattime
      \dotplot{red}{\nsattime};
      \addplot [name path=upper,draw=none]
      table[x=x,y expr=\thisrow{y}+\thisrow{err}] {\nsattime};
      \addplot [name path=lower,draw=none] 
      table[x=x,y expr=\thisrow{y}-\thisrow{err}] {\nsattime};
      \linedevplot{red}
  \end{axis}
    \end{tikzpicture}
    \caption{
      Left axis: the total number of calls to the \sat-solver (sat-count).
      Right axis: the total time spent waiting on the \sat-solver (sat-time).
    }\label{fig:peter:relax:4proc:sat}
  \end{subfigure}
  \caption{
    Relaxing \ttfunction{Peterson4} experiment (10 repetitions). Left axis: the number
    of counterexamples (cti-count) found in the major iteration. Right axis:
    the total runtime (cti-time).
  }
  \label{fig:peter:relax:4proc:oblsat}
\end{figure}
}

\end{document}